\PassOptionsToPackage{dvipsnames}{xcolor}
\documentclass[twocolumn, nolinenumbers]{aastex631}
\watermark{}

\usepackage{acronym}

\usepackage{xspace}
\usepackage[dvipsnames]{xcolor}
\usepackage[normalem]{ulem}
\usepackage{amsmath}
\usepackage{ragged2e}
\usepackage{booktabs}
\usepackage{multirow}
\usepackage{paralist}
\usepackage{hyperref}
\usepackage{booktabs}
\usepackage{multirow}
\usepackage{soul}

\PassOptionsToPackage{breaklinks, plainpages=false, anchorcolor=cyan, linkcolor=magenta, citecolor=cyan, urlcolor=violet, bookmarks=false}{hyperref}

\newacro{imbh}[IMBH]{intermediate-mass black hole}
\newacro{bhns}[BHNS]{black hole neutron star}
\newacro{bbh}[BBH]{binary black hole}
\newacro{bh}[BH]{black hole}
\newacro{bns}[BNS]{binary neutron star}
\acrodef{FAR}[FAR]{false alarm rate}
\newacro{bf}[BF]{Bayes' factor}
\newacro{cbc}[CBC]{compact binary coalescence}
\newacro{ce}[CE]{Cosmic Explorer}
\acrodef{SNe}[SNe]{Supernova}
\newacro{da}[DA]{data analysis}
\newacro{et}[ET]{Einstein Telescope}
\newacro{eob}[EOB]{Effective-One-Body}
\newacro{fd}[FD]{frequency domain}
\newacro{gw}[GW]{gravitational-wave}
\newacro{gr}[GR]{general relativity}
\newacro{hm}[HM]{Higher mode}
\newacro{ifo}[IFO]{interferometer}
\newacro{imr}[IMR]{inspiral-merger-ringdown}
\newacro{im}[IM]{inspiral-to-merger}
\newacro{kagra}[KAGRA]{Kamioka Gravitational Wave Detector}
\newacro{ligo}[LIGO]{Laser Interferometer Gravitational-Wave Observatory}
\newacro{lso}[LSO]{Last Stable Orbit}
\newacro{lvc}[LVC]{LIGO-Virgo Collaboration}
\newacro{lvk}[LVK]{LIGO-Virgo-Kagra Collaboration}
\newacro{lo}[LO]{leading order}
\newacro{ns}[NS]{neutron star}
\newacro{nr}[NR]{numerical relativity}
\newacro{pn}[PN]{post-Newtonian}
\newacro{pe}[PE]{parameter estimation}
\newacro{psd}[PSD]{power spectral density}
\newacro{asd}[ASD]{amplitude spectral density}
\acrodef{kn}[KN]{kilonova}
\newacro{xg}[XG]{next-generation}
\newacro{jsd}[JSD]{jensen shannon divergence}
\newacro{sgrb}[sGRB]{short gamma ray burst}
\newacro{igwn}[IGWN]{international gravitational wave network}
\newacro{qc}[QC]{quasi-circular}
\newacro{snr}[SNR]{signal-to-noise ratio}
\newacro{eos}[EoS]{equation of state}
\newacro{em}[EM]{electromagnetic}

\newcommand{\gds}{golden dark sirens}
\newcommand{\gd}{golden dark siren}
\newcommand{\nrhyb}{\texttt{NRHybSur3dq8}}
\newcommand{\seob}{\texttt{SEOBNRv5PHM}}
\newcommand{\teob}{\texttt{TEOBResumS-GIOTTO}}
\newcommand{\imrxpnr}{\texttt{IMRPhenomXPNR}}
\newcommand{\imrst}{\texttt{IMRPhenomXPHM-SpinTaylor}}
\newcommand{\BILBY}{\texttt{BILBY}}
\newcommand{\comment}[1]{}

\begin{document}
% \onehalfspacing

% \preprint{APS/123-QED}

\title{Tarnished by Tools: Cost of Systematics in Golden Dark Siren Cosmology}

\author{Giovanni Benetti}
    \email{giovanni.benetti.1@studenti.unipd.it}
    \affiliation{Dipartimento di Fisica e Astronomia Galileo Galilei, Università di Padova, 35131 Padova, Italy}
    \affiliation{Institute for Gravitation \& the Cosmos, Department of Physics, Penn State University, University Park, PA 16802, USA} 
    \thanks{G.B. and K.C. contributed equally to this work.}
\author{Koustav Chandra}
    \email{koustav.chandra@aei.mpg.de}
    \affiliation{Max Planck Institute for Gravitational Physics (Albert Einstein Institute), Am M\"uhlenberg 1, 14476 Potsdam, Germany}
    \affiliation{Institute for Gravitation \& the Cosmos, Department of Physics, Penn State University, University Park, PA 16802, USA}
    \affiliation{Department of Astronomy and Astrophysics, Penn State University, University Park, PA 16802, USA}
    \thanks{G.B. and K.C. contributed equally to this work.}

\author{Bangalore S. Sathyaprakash}
    \affiliation{Institute for Gravitation \& the Cosmos, Department of Physics, Penn State University, University Park, PA 16802, USA}
    \affiliation{Department of Astronomy and Astrophysics, Penn State University, University Park, PA 16802, USA}

\date{\today}

\begin{abstract}
Golden dark sirens—exceptionally well-localized gravitational-wave (GW) sources without electromagnetic counterparts—offer a powerful route to precision measurements of the Hubble constant, $H_0$, with next-generation (XG) detectors. The statistical promise of this method, however, places stringent demands on waveform accuracy and detector calibration, as even small systematic errors can dominate over statistical uncertainties at high signal-to-noise ratios. We investigate the impact of waveform-modeling systematics on golden dark siren cosmology using a synthetic population of binary black holes consistent with current GW observations and analyzed in the XG-detector era. By comparing state-of-the-art waveform models against numerical-relativity–based reference signals, we quantify modeling inaccuracies from both modeling and data-analysis perspectives and assess how they propagate into biases in luminosity distance, host-galaxy association, and single-event $H_0$ inference. We find that while current waveform models often allow recovery of statistically consistent $H_0$ posteriors, small waveform-induced biases can significantly affect three-dimensional localization and host galaxy ranking, occasionally leading to incorrect redshift assignments. We further derive order-of-magnitude requirements on detector calibration accuracy needed to ensure that calibration systematics remain subdominant for golden dark sirens observed with XG networks. To realize sub-percent $H_0$ measurements with golden dark sirens will require waveform and calibration accuracies that scale as $\mathcal{O}(\rho^{-2})$ with signal-to-noise ratio, motivating sustained advances in waveform modeling, numerical relativity, and detector calibration for the XG era.
\end{abstract}

%%%%%%%%%%%%%%%%%%%%%%%%%%%%%%%%%%%%%%%%%%%%%%%%%%%%%%%%%
\section{Introduction}
%%%%%%%%%%%%%%%%%%%%%%%%%%%%%%%%%%%%%%%%%%%%%%%%%%%%%%%%%

\Acp{gw}, like light, get redshifted as they propagate through the expanding Universe, and therefore bear information about the cosmic expansion. However, unlike light, each \ac{gw} event provides a direct measurement of the luminosity distance \(D_L\) to its source, independent of the cosmic distance ladder. When combined with an independent redshift \((z)\) measurement, the resulting ``standard siren'' enables direct inference of cosmological parameters, such as the Hubble constant \(H_0\)~\citep{Schutz:1986gp, Holz:2005df}.

A textbook example of such a standard siren is the multi-messenger event GW170817 \citep{LIGOScientific:2017vwq} by the Laser Interferometer Gravitational-Wave Observatory (LIGO) \citep{LIGOScientific:2014pky} and Virgo \citep{VIRGO:2014yos}, whose unambiguous association with its host galaxy NGC 4993 provided a precise redshift measurement, thereby allowing for the first direct ``bright siren'' measurement of \(H_0\)~\citep{LIGOScientific:2017adf}. Yet, the low merger rate of \ac{bns} systems means that GW170817 remains the only reliable bright siren detected over four observing runs of advanced ground-based \ac{gw} interferometers~\citep{LIGOScientific:2025pvj}. Few studies have claimed electromagnetic associations with \ac{bbh} mergers and used them to report additional constraints/estimates, but these counterparts are highly contentious~\citep{Graham:2020gwr, Ashton:2020kyr, Gayathri:2020mra, Mukherjee:2020kki, Leong:2025qiw}; hence, the reported estimates are unreliable.

{These limitations have prompted the development and usage of alternative strategies for \ac{gw} cosmology (within the \ac{gr}-consistent framework) that do not require a prompt electromagnetic counterpart. These include:
\begin{inparaenum}[(1)]
\item leveraging galaxy catalogs to obtain an ensemble of probabilistic ``dark siren'' measurements of \(H_0\) by combining multiple \ac{gw} events~\citep{DelPozzo:2011vcw, Chen:2017rfc, LIGOScientific:2018gmd, Gray:2019ksv, Gray:2021sew, Leandro:2021qlc, Gair:2022zsa, Chen:2024gdn};
\item cross-correlating the sky distribution of \ac{gw} sources with galaxies—again relying on galaxy catalogs—to extract redshift information statistically through their spatial clustering~\citep{Oguri:2016dgk, Mukherjee:2019wcg, Mastrogiovanni:2021wsd, Bera:2020jhx,  Mukherjee:2022afz, LIGOScientific:2025jau};
\item performing ``spectral-siren'' inference, in which the redshift is constrained directly from the population properties of the \ac{gw} sources~\citep{Chernoff:1993th, Taylor:2011fs, Farr:2019twy, Mukherjee:2021rtw, Karathanasis:2022rtr}; and
\item using the tidal deformabilities of binary components together with their detector-frame (redshifted) masses to infer the redshift~\citep{Messenger:2011gi, Ghosh:2022muc, Ghosh:2024cwc}.
\end{inparaenum}}

The first alternative approach, the dark-siren \(+\) galaxy catalog method, uses redshift information from all potential host galaxies within the three-dimensional localization volume of a \ac{gw} event, and has been employed since the discovery of GW170814 — the first three-detector \ac{bbh} observation~\citep{LIGOScientific:2017ycc, DES:2019ccw}. Since the localization volume decreases sharply with increasing \ac{snr}, exceptionally loud events can be constrained to regions containing only a single plausible host galaxy. Such sources, informally termed \textit{\gds}, enable unique host identification without an electromagnetic counterpart. Following the working definition of \citet{Borhanian:2020vyr}, non-electromagnetically bright events at $z \leq 0.1$ confined to a sky area smaller than $0.04~\mathrm{deg}^2$ are expected to contain only one $L^\ast$ galaxy within their localization region and thus qualify as \gds.

Forecasts for \ac{xg} detectors highlight the promise of \gds~cosmology. \citet{Borhanian:2020vyr} estimate that even a network upgraded to the \(A^+\) sensitivity level could achieve an \(\mathcal{O}(1\%)\) measurement of \(H_0\) at 68\% credibility using \gds~alone. A more advanced network comprising two \acp{ce} (with 40\,km and 20\,km arms) together with a triangular \ac{et} (with 10 km arms) is projected to reach \(\mathcal{O}(0.1\%)\) precision within two years of operation, assuming \ac{bbh} population estimates of \(\sim 10\) mergers per year at \(z \lesssim 0.1\) in \ac{xg} era (also see \cite{Chen:2024gdn, Gupta:2023lga}).

However, such high statistical precision places stringent demands on all aspects of the analysis, from detector calibration to noise characterization \citep{Vitale:2011wu, Mozzon:2021wam, Essick:2022vzl, Capote:2024mqe}. In particular, waveform accuracy becomes particularly critical~\citep{Purrer:2019jcp, Chandra:2024dhf, Dhani:2024jja}, as even small waveform inaccuracies can bias the inferred redshift–distance relation and, consequently, the \(H_0\) measurement. Moreover, \(H_0\) is not an observable of any single event but a \textit{hyperparameter} that is expected to be consistent across an ensemble of sources. Broader investigations of the impact of systematics on standard sirens cosmology have already highlighted the challenges. For example, \citet{Dhani:2024jja} showed that waveform systematics can significantly bias gravitational-wave inferences of luminosity distance and sky localization, even for high–signal-to-noise events in advanced and \ac{xg} detector networks, with the largest effects arising for mass-asymmetric and spin-precessing binaries; such biases can lead to incorrect host-galaxy association and biased cosmological inference, including of the Hubble constant, despite excellent localization. 

Following up on this, \citet{Dhani:2025xgt} showed that, for current detector networks, a small number of \ac{bbh} events with strong mass asymmetry or spin precession can disproportionately bias dark-siren $H_0$ inference, while future facilities will infer a biased $H_0$ value for even the current-empirically inferred \ac{bbh} population, when using state‑of‑the‑art waveform models within a hierarchical analysis. Similarly, \citet{Hanselman:2024hqy} showed that host‑galaxy weighting itself can bias dark‑siren cosmology: incorrect assumptions about the galaxy redshift distribution or weighting prescription can systematically favor the wrong hosts. The severity of this bias depends on the number of galaxies along the line of sight and the uncertainty in the luminosity‑distance measurement. They further demonstrated that improved localization and better‑informed priors can mitigate these effects. 

Both studies advocate using a hierarchical framework to identify and mitigate biases throughout the analysis consistently. In the context of bright sirens with current detectors, \citet{Huang:2022rdg} found that sufficiently large calibration errors across the detector network can bias joint inferences of $H_0$, particularly if a significant fraction of the observed population suffers from the same large bias. Taken together, these results underscore that the constraining power of \gds{} cosmology relies not only on improved detector sensitivity, but also on the robustness of waveform models, calibration, and population inference. Without commensurate control of systematic uncertainties, modeling errors may ultimately limit the precision achievable in \(H_0\) measurements with \ac{xg} detectors.

Here, we investigate how systematics affect \gds{} cosmology in the era of \ac{xg} detectors. Using a synthetic GWTC-3–inspired \ac{bbh} population and focusing on the most informative events—those whose exceptional localization enables unique host-galaxy identification—we quantify how inaccuracies in state-of-the-art waveform models bias the inferred sky location, luminosity distances and, consequently, the recovered value of \(H_0\). We find that while the true host galaxy is correctly identified for most events, waveform-induced biases in \(D_L\) can, in some cases, shift the preferred host, thereby yielding erroneous \(z\) assignments and biased \(H_0\) measurements.

%%%%%%%%%%%%%%%%%%%%%%%%%%%%%%%%%%%%%%%%%
\section{Waveform models comparison}

\label{sec:wsys}
%%%%%%%%%%%%%%%%%%%%%%%%%%%%%%%%%%%%%%%%%

\begin{figure*}
    \centering
    \includegraphics[width=0.98\textwidth]{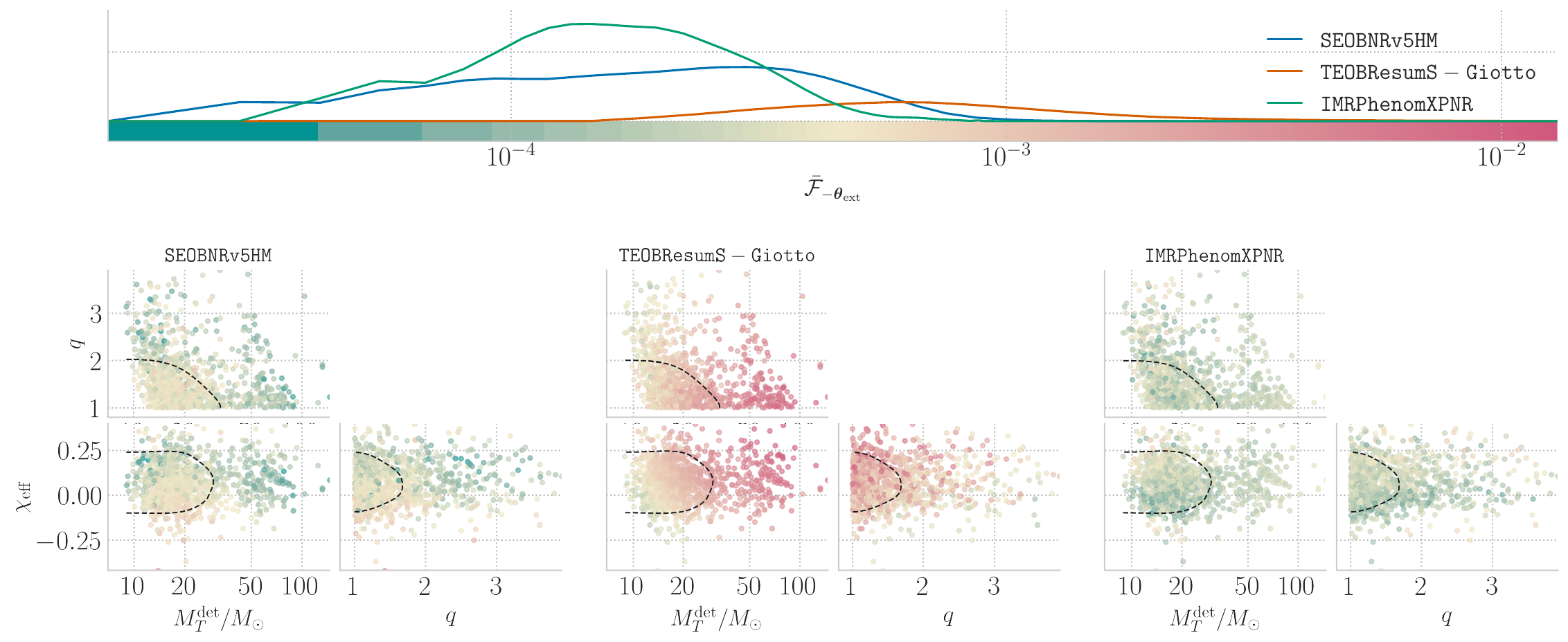}
    \caption{Waveform unfaithfulness for a simulated population of golden dark sirens, evaluated from a modeller's perspective. \textit{Top panel:} Distribution of extrinsic-parameter–maximized unfaithfulness, \(\bar{\mathcal{F}}_{\boldsymbol{\theta}_{\rm ext}}\) for three different waveform models \seob{} (blue), \teob{} (red), and \imrxpnr{} (green)—computed against the reference signal population simulated using \nrhyb. \textit{Bottom panel:} Event-by-event unfaithfulness projected across the intrinsic parameter space, shown as functions of detector-frame total mass \(M_T^{\rm det}\), mass ratio \(q = m_1/m_2 \geq 1\) and effective inspiral spin parameter \(\chi_{\rm eff}\). Each column represents a different waveform family. Colors indicate the value of \(\bar{\mathcal{F}}_{\boldsymbol{\theta}_{\rm ext}}\), highlighting regions of parameter space where modelling systematics are most pronounced. The black-dashed lines encompass \(68\%\) of the simulated population.}
    \label{fig:unfaithfulness-1}
\end{figure*}

Accurate inference of \(H_0\) from \gds{} relies critically on the fidelity of the waveform models used in single-event parameter estimation. In the high-\ac{snr} regime relevant for \ac{xg} detectors, even small modelling inaccuracies can lead to biased parameter recovery, with potential consequences for host-galaxy association and, ultimately, population-level cosmological constraints. As a first step, we here use faithfulness \(\mathcal{F}\) defined as~\citep{Apostolatos:1995pj},
\begin{equation}\label{eq:fitting-factor}
    \mathcal{F}
    = \max_{\boldsymbol{\Lambda}}
    \frac{ \left( s \mid h(\boldsymbol{\theta}) \right) }
         { \sqrt{ \left( s \mid s \right) \left( h \mid h \right) } }\,,
\end{equation}
and the unfaithfulness \(\bar{\mathcal{F}} = 1-\mathcal{F}\) to quantify the agreement between two waveforms after maximization over a chosen set of parameters \(\boldsymbol{\Lambda} \subset \boldsymbol{\theta}\). Origin of bias due to waveform systematics and its relation to \(\bar{\mathcal{F}}\) is summarized in Appendix~\ref{app:origin}.

We generate multiple realizations of a synthetic \ac{bbh} population with hyperparameters consistent with GWTC--3 (see Appendix~\ref{sec:configuration}). Most simulated ``golden'' binaries have near-equal masses, low spins, and small spin--orbit misalignment—precisely the regime where current waveform models are expected to perform best. Since very few \ac{gr}-faithful \ac{nr} simulations are long enough to cover the full \ac{xg} bandwidth, we adopt \nrhyb{} waveforms as our reference signals \(s\)~\citep{Varma:2018mmi, Yoo:2023spi}~\footnote{Here, we use the model version which doesn't contain \ac{gw} memory effects.}. \nrhyb{} is a hybrid \ac{nr} surrogate model built by training on \ac{nr}--\ac{pn}/\ac{eob} hybrid waveforms for quasi-circular systems with mass ratios \(q\leq 8\) and spin magnitudes \(\chi_{1,2}\leq 0.8\)~\footnote{Throughout, ``quasi-circular'' denotes non-eccentric, non-precessing binaries.}. This choice is well-suited to our source population, which exhibits small in-plane spins and no large mass-asymmetry --- \emph{in other words, we assume our source population is quasi-circular and within the training range of the surrogate model}. For a white noise spectrum, we note that the unfaithfulness between \nrhyb{} and the \ac{nr} part of the training hybrid waveforms has a median \(\bar{\mathcal{F}} \approx 10^{-5}\). The \ac{nr} waveforms used for the training have a self-\(\bar{\mathcal{F}} \geq 10^{-7}\) for a design-sensitivity \ac{ce} noise curve~\citep{Scheel:2025jct}.

We compare three state-of-the-art waveform models—\seob{}~\citep{Pompili:2023tna, Ramos-Buades:2023ehm}, \teob{}~\citep{Riemenschneider:2021ppj, Nagar:2023zxh}, and \imrxpnr{}~\citep{Hamilton:2025xru}—against \nrhyb{}. The \seob{} and \teob{} waveform models compute the multipolar radiation within the \ac{eob} framework~\citep{Buonanno:1998gg, Buonanno:2000ef}, matching a resummed \ac{pn}-based inspiral--plunge waveform calibrated to \ac{nr} to a merger--ringdown model informed by black-hole perturbation theory and \ac{nr} simulations. By contrast, \imrxpnr{} is a closed-form, frequency-domain phenomenological model constructed across the inspiral, merger, and ringdown phases of the binary evolution using augmented \ac{pn} expressions and calibrated polynomial and Lorentzian fits. Built on the \texttt{IMRPhenomXO4a} framework, it incorporates \imrst{} precession dynamics and \ac{nr}-informed multipole asymmetries, thereby providing the most accurate phenomenological description currently available for generic, non-eccentric \ac{bbh} signals~\citep{Pratten:2020ceb, Thompson:2023ase, Colleoni:2024knd}.

%%%%%%%%%%%%%%%%%%%%%%%%%%%%%%%%%%%%%%%%%
\section{Modeller's perspective}
%%%%%%%%%%%%%%%%%%%%%%%%%%%%%%%%%%%%%%%%%

A waveform modeller hopes to recover the \emph{true} physical parameters of a \ac{gw} signal, limited only by detector noise. Therefore, when assessing waveform-modelling systematics, the comparison must be carried out using \emph{identical intrinsic parameters} for both the signal and the template waveform. This isolates discrepancies in the waveform morphology—i.e., the modelling error. Under this restriction, the maximization in Eq.~\eqref{eq:fitting-factor} reduces to the subspace of extrinsic parameters. Practically, this involves \textit{effectively extremizing} over the luminosity distance, sky location, and polarization angle, following, e.g., Eq.~(13) of~\citet{Harry:2017weg} or Eq.~(10) of~\citet{Chandra:2022ixv}. Additionally, one maximizes over the signal's peak time by locating the peak of the faithfulness time series via inverse Fourier transforms, and over the binary's azimuth to account for different conventions across waveform models. Since the inclination angle controls the prominence of higher-order multipoles, the statistic is evaluated over a range of inclinations, and the minimum unfaithfulness \(\bar{\mathcal{F}} = 1 - \mathcal{F}\) is reported. Likewise, the effective polarization angle \(\kappa\) is varied, and the resulting unfaithfulness is averaged over this degree of freedom. Together, the extremization is restricted to the extrinsic parameter vector \(\boldsymbol{\theta}_{\rm ext}\), and the corresponding tolerance (se App. \ref{app:origin}) may be written as \(\epsilon = N\), where \(N\) \textit{denotes the number of binary parameters that are not maximized over}~\citep{Chatziioannou:2017tdw}. As a result, any discrepancy thus obtained reflects differences in waveform morphology rather than underlying source description. Henceforth, we will use $\bar{\mathcal{F}}_{-\boldsymbol{\theta}_{\rm ext}}$ for the unfaithfulness maximized over extrinsic parameters only, and $\bar{\mathcal{F}}$ for the unfaithfulness maximized over all waveform parameters.

Fig.~\ref{fig:unfaithfulness-1} shows the \(\bar{\mathcal{F}}_{-\boldsymbol{\theta}_{\rm ext}}\) distribution and its dependence on the intrinsic parameters of a synthetic \gd{} signal population simulated using \nrhyb{}. We consider only the radiation multipoles \((\ell,|m|)=((2,2),(3,3),(2,1), (4,4))\) for all the waveform models to avoid discrepancies arising from different subdominant mode content. The top panel compares the \(\bar{\mathcal{F}}_{-\boldsymbol{\theta}_{\rm ext}}\) distribution for the three state-of-the-art waveform families. Interpreted using Eq.~\eqref{eq:criterion} and setting \(N=4\), we find that only \(\sim 2\%\) of the population analysed with \imrxpnr{} satisfies \(\bar{\mathcal{F}}_{-\boldsymbol{\theta}_{\rm ext}} \leq 5\times10^{-5}\); this fraction drops to \(\lesssim 1.3\%\) for \seob{}, and to zero for \teob{}~\footnote{We note that this number doesn't change even when we use the other \texttt{TEOBResumS} avatar, \texttt{TEOBResumS-Dalí}. The \(\bar{\mathcal{F}}_{-\boldsymbol{\theta}_{\rm ext}}\) distribution remains almost identical with Jensen-Shannon divergence of \(\approx 0.05\) nats.}. In other words, for signals with \(\rho \gtrsim 200\), \textit{waveform systematics may dominate statistical uncertainties}, leading to biased parameter inferences.

We note that the waveform agreement is expected. The \nrhyb{} waveform model and \imrxpnr{} (built upon \imrst{}) are calibrated to \texttt{SEOBNRv4HM} \citep{Cotesta:2018fcv}, whereas \seob{} improves upon its fourth version. Consistent with this calibration history, \imrxpnr{} exhibits greater agreement with \nrhyb{}.

The lower panels project the same unfaithfulness values onto the intrinsic parameter space spanned by the detector-frame total mass \(M_T^{\rm det}\), mass ratio \(q = m_1/m_2 \geq 1\), and effective inspiral spin \(\chi_{\rm eff}\). The dashed contours enclose \(68\%\) of the simulated population and demonstrate that appreciable waveform disagreement persists even for nearly equal-mass, low-spin binaries. At low total masses, where the signal accumulates more inspiral cycles within the detector bandwidth, \imrxpnr{} yields smaller values of \(\bar{\mathcal{F}}_{-\boldsymbol{\theta}_{\rm ext}}\) than \seob{}, indicating better phase coherence with the inspiral-portion (\ac{pn} portion) of the \nrhyb{} waveforms. At higher masses (\(M_T^{\rm det} \gtrsim 50\,M_\odot\)), \seob{} performs marginally better, consistent with its better \ac{nr} calibration of post-inspiral phase. \teob's performance worsens as total mass increases, reflecting a comparatively poorer agreement with the reference signals.

%%%%%%%%%%%%%%%%%%%%%%%%%%%%%%%%%%%%%%%%%
\section{Data analyst's perspective}
%%%%%%%%%%%%%%%%%%%%%%%%%%%%%%%%%%%%%%%%%
From a data analyst's viewpoint, disagreements between the signal and the template waveforms for identical intrinsic parameters are not necessarily as problematic as they might first appear. In fact, stochastic samplers are designed to map out the posterior distribution, not to ``home in'' on the best-fit parameters. Consequently, waveform modelling errors need not bias inference, provided that the template waveform reproduces the signal at a \emph{different} set of parameters that lies within a predetermined credible region.

Now, performing a full-scale Bayesian parameter estimation to confirm this for every event in our synthetic population is computationally prohibitive. Therefore, as a first step, we adopt an approximate diagnostic. We assume that the posteriors are well-approximated by a multivariate Gaussian in \(N\) parameters, with the best-fit parameters lying near the posterior's peak. For this assumption, the drop in maximum log-likelihood from its intended maximum due to systematics,
\(2\Delta \ln \mathcal{L} \equiv 2(\ln \hat{\mathcal{L}} (\boldsymbol{\lambda})-\ln \mathcal{L}_{\rm max} (\boldsymbol{\theta}))\),
is approximately distributed as \(\chi^2_N\). Then the boundary containing the \(90\%\) of the posteriors is given by:
\begin{equation}\label{eq:criterion-2}
    2\Delta \ln \mathcal{L}
    \simeq \rho^2 \bar{\mathcal{F}}
    < \mathcal{Q}_{0.9}(N) \implies \bar{\mathcal{F}} < \frac{\mathcal{Q}_{0.9}(N)}{\rho^2}
\end{equation}
where \(\mathcal{Q}_{0.9}(N)\) denotes the \(90\%\) quantile of the \(\chi^2_N\) distribution. Geometrically, this inequality demands that parameter shifts induced by waveform-modelling errors remain confined within the noise-dominated posterior ellipsoid at the chosen confidence level. This is identical to the indistinguishability criterion of \citet{Toubiana:2024car}, who use the Wilson–Hilferty transformation to go one step further. However, there are caveats associated with this inequality, which we elaborate on later.

\begin{figure}
    \centering
    \includegraphics[width=0.5\textwidth]{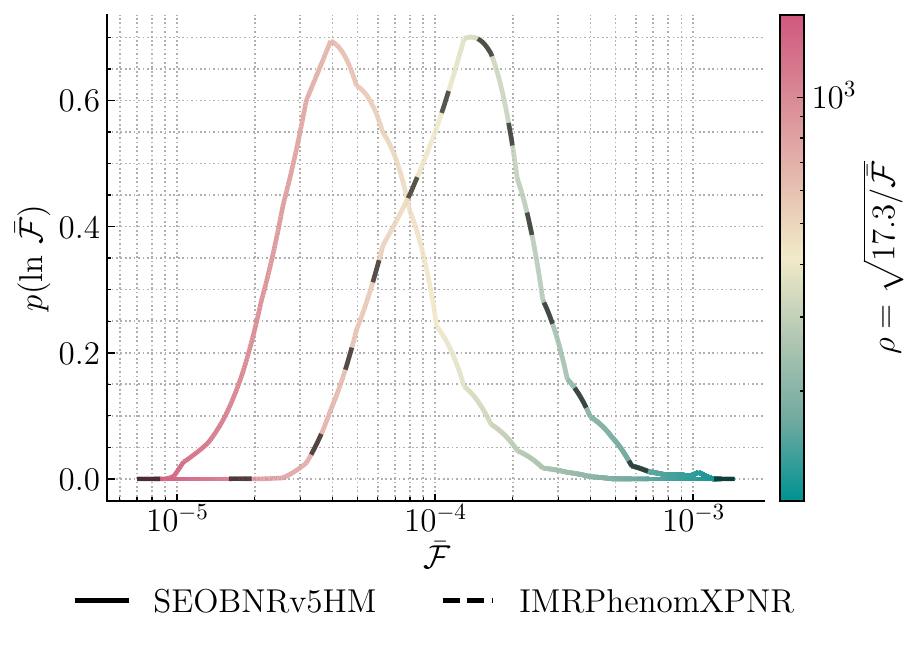}
    \caption{Waveform unfaithfulness for a simulated population of \gds{}, evaluated from a data‑analysis perspective. The solid curve shows the distribution of \(\bar{\mathcal{F}}\) for \seob{} when compared against \nrhyb{}, while the curve with overlaid black dashed lines on top shows the corresponding distribution for \imrxpnr{}. The color scale indicates the \ac{snr} at which waveform systematics would fail to recover the true binary parameters within their 90\% credible level.}
    \label{fig:data-analyst}
\end{figure}

Using the above equation, we compare $\sim 1000$ \gds{} signals to \imrxpnr{} and \seob{} template models, assuming a three-detector \ac{xg} network operating at design sensitivity. For each signal, the fitting factor in Eq.~\eqref{eq:fitting-factor} is maximized over all binary parameters using a hybrid strategy involving differential evolution and Nelder-Mead~\citep{Storn:1997uea, Gao:2012guu, 2020SciPy-NMeth}. This procedure is equivalent to recovering the signal in the absence of noise while focusing on just the best-fitting waveform. Given the caveats involved, we repeat the optimization five times with different random seeds and retain the solution yielding the smallest value of \(\bar{\mathcal{F}}\). We do not perform this analysis with \teob{} owing to its comparatively higher \(\bar{\mathcal{F}}_{-\boldsymbol{\theta}_{\rm ext}}\). Also, given that we perform the optimization over all quasi-circular binary parameters, we set \(N = 11\) for which Eq.~\eqref{eq:criterion-2} reduces to $\bar{\mathcal{F}} < 17.3/\rho^2$; thus, for $\rho=300$, the requirement becomes $\bar{\mathcal{F}} \lesssim 2\times 10^{-4}$ for waveforms to be indistinguishable.

Fig.~\ref{fig:data-analyst} shows the resulting distribution of \(\bar{\mathcal{F}}\), color-coded by the network \ac{snr} at which waveform-modelling systematics are expected to become comparable to statistical uncertainties. The solid line corresponds to \nrhyb-\seob{} case while the curve with overlaid black dashed line indicates the \nrhyb{}-\imrxpnr{} case. This diagnostic indicates that \seob{}, when allowed to adjust its intrinsic parameters to represent the signal adequately, can, in principle, obtain lower unfaithfulness.

\begin{figure*}
    \centering
    \includegraphics[width=1\linewidth]{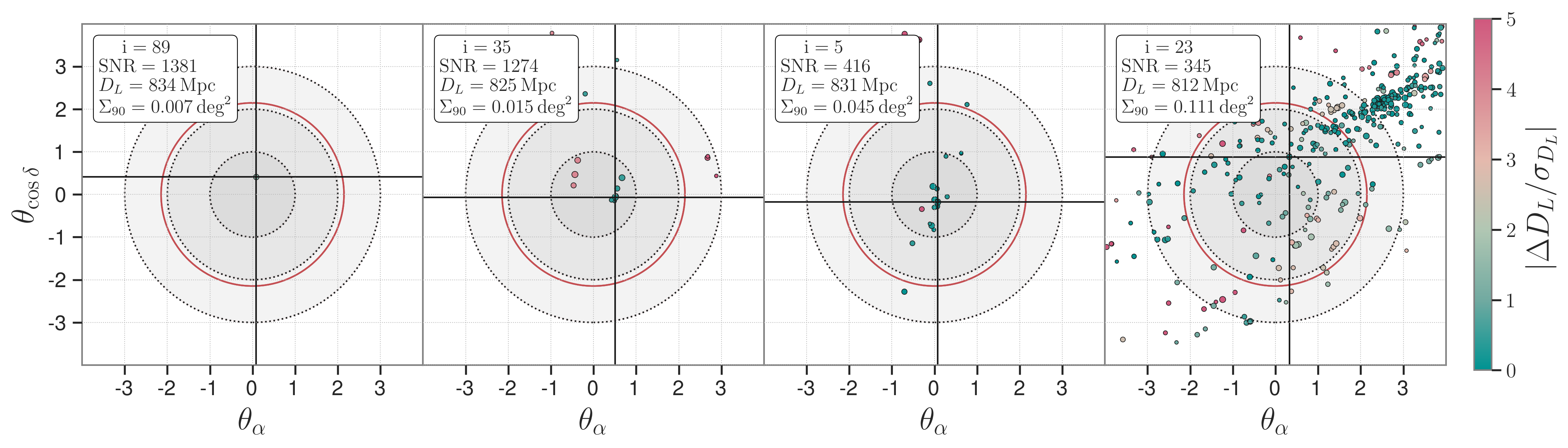}
    \caption{Representative whitened sky localizations for \imrxpnr{} recoveries at comparable luminosity distance $D_L$. The true host galaxy is marked by intersecting black lines, while other candidate host galaxies are color-coded by their luminosity-distance offset relative to the inferred $D_L$, in units of the posterior standard deviation $\sigma_{d_L}$. Grey contours indicate the $|\boldsymbol{\Theta}|=1\sigma,\,2\sigma,$ and $3\sigma$ regions, and the red circle encloses the 90\% credible sky area $\Sigma_{90}$. From left to right, the panels illustrate a golden siren with a unique host, two copper sirens with \(<10\) plausible hosts at similar distances, and a classical dark siren with several viable counterparts.}
    \label{fig:whitened_skyloc_selected}
\end{figure*}

We emphasize, however, that these results should be interpreted with care.
\begin{inparaenum}[(1)]
    \item First, achieving a lower \(\bar{\mathcal{F}}\) by adjusting waveform parameters does not necessarily indicate an improved physical description of the signal: such adjustments can instead compensate for deficiencies in the waveform model by absorbing modelling differences into biased parameter values, leading to erroneous inference.
    \item  Second, although we have tried to mitigate optimization effects by repeating the maximization with multiple random seeds and validating with dual-annealing, residual numerical and algorithmic limitations—especially in a highly structured, high-dimensional likelihood—may still affect the recovered extrema at such high numerical accuracy. 
\end{inparaenum}
The quoted fractions should therefore be viewed as indicative of the onset of systematics rather than precise predictions of parameter biases, which we attempt to quantify next using Bayesian inference on a representative subset of signals.

%%%%%%%%%%%%%%%%%%%%%%%%%%%%%%%%%%%%%%%%%
\section{Small Biases, Big Consequences}
\label{sec:bayesian}
%%%%%%%%%%%%%%%%%%%%%%%%%%%%%%%%%%%%%%%%%

We now examine how waveform-induced biases propagate to \(H_0\) measurement by focusing first on their impact on the inferred three‑dimensional source localization—sky position \((\alpha, \delta)\) and luminosity distance \(D_L\). We construct a mock catalog of $111$ \acp{bbh} detectable at redshifts $z\lesssim0.1$ by a three-detector \ac{xg} network over its first ten years of operation, assuming a duty cycle of 0.8 and a source population consistent with the GWTC-3 inference. Each event is assigned a three-dimensional localization by associating it with galaxies drawn from the MICECATv2.0 catalog~\citep{Fosalba:2013wxa, Fosalba:2013mra, Crocce:2013vda, Carretero:2014ltj, Hoffmann:2014ida}, with host probabilities weighted by galaxy mass.

Signals from these binaries are simulated using \nrhyb{} and analyzed in the absence of noise using the waveform models \imrxpnr{} and \seob{}, employing the parameter-estimation package \BILBY{} without likelihood-acceleration techniques~\citep{Ashton:2018jfp}. Unless otherwise stated, we use the nested sampler \texttt{dynesty} with settings typical of current \ac{gw} analyses \((n_{\mathrm{live}}=10^3,~n_{\mathrm{accept}}=60, ~\mathtt{maxmcmc}=5 \times 10^3)\)~\citep{Speagle:2019ivv}. For a representative subset of high-\ac{snr} events, we repeat the analysis with more aggressive configurations \((n_{\mathrm{live}}\gtrsim4 \times 10^3,~ \texttt{maxmcmc}=10^4)\), obtaining consistent posteriors and confirming that waveform-modelling systematics rather than inference limitations dominate the results presented below.
    
%%%%%%%%%%%%%%%%%%%%%%%%%%%%%%%%%%%%
\section{3D Source localization}
%%%%%%%%%%%%%%%%%%%%%%%%%%%%%%%%%%%%
\label{sec:3DSourcelocalization}
We begin by characterising biases in sky position and luminosity distance separately, before turning to their combined impact on three-dimensional localization. To enable a uniform comparison across events with widely varying posterior widths, we whiten the sky-location coordinates $\boldsymbol{\Omega}=(\alpha,\cos\delta)$. For the high-\ac{snr} signals considered here, sky-localization posteriors are well approximated by Gaussians, since the likelihood is dominated by its quadratic expansion about the maximum. Under this assumption, the joint sky posterior may be written as
\begin{equation}
    P(\boldsymbol{\Omega}) \propto 
    \exp\!\left[-\frac{1}{2}
    (\boldsymbol{\Omega}-\boldsymbol{\Omega}_b)\,
    \mathcal{C}_{\mathrm{sky}}^{-1}\,
    (\boldsymbol{\Omega}-\boldsymbol{\Omega}_b)^{T}
    \right],
\end{equation}
where $\boldsymbol{\Omega}_b$ denotes the best-fit sky location and $\mathcal{C}_{\mathrm{sky}}$ is the corresponding covariance matrix, obtained from the posterior samples. We perform a linear transformation to whitened coordinates $\boldsymbol{\Theta}=(\theta_\alpha,\theta_\delta)$ such that the transformed posterior is a two-dimensional Gaussian with zero mean and unit covariance, $\boldsymbol{\Theta}_b=0$ and $\mathcal{C}_{\rm sky}^{ij}=\delta^{ij}$. This whitening is implemented using zero-phase component analysis (ZCA), as described in Appendix~\ref{app:whitening}.

Fig.~\ref{fig:whitened_skyloc_selected} shows representative examples of \imrxpnr{} recovery at comparable luminosity distance in the whitened coordinate plane. The true host galaxy is marked by intersecting red lines, while other candidates are color-coded by their offsets in luminosity distance relative to the inferred value, expressed in units of the posterior standard deviation $\sigma_{D_L}$. Grey contours enclose regions of $|\boldsymbol{\Theta}| = 1\sigma$, $2\sigma$, and $3\sigma$, while the red circle denotes the $90\%$ sky area.

The leftmost panel illustrates a case (system~89) that qualifies as a golden dark siren, with a uniquely and correctly identified host galaxy. The middle panels (systems~5 and~35) correspond to ``copper sirens,'' where multiple but \(<10\) plausible hosts lie within the \(90\%\) sky region and have luminosity distances consistent with that of the true host. The rightmost panel shows an event with several viable host candidates, representative of a classical dark siren for which host association is more naturally addressed through galaxy-catalog marginalization. The difference in galaxy density around the source is primarily driven by the sky localization precision (note the SNR and 90\% sky area progression between the systems) and by anisotropies in the whitened coordinates, caused by squeezing along the major axis of the physical sky location posterior.

To quantify the impact of waveform-modelling systematics on host-galaxy association for the full \gds{} population (see Appendix~\ref{sec:3d}), we consider only galaxies within the 90\% sky credible region whose redshifts are consistent with the inferred luminosity distance under the assumption of a flat \(\Lambda\)CDM cosmology with matter and energy densities fixed to \citet{Planck:2018vyg} and a flat prior on $H_0 \in (60,75)\,\mathrm{km\,s^{-1}\,Mpc^{-1}}$. For \imrxpnr{} recovery, approximately $80\%$ ($85\%$, $95\%$) of true host galaxies are contained within the $50\%$ ($67\%$, $90\%$) three-dimensional credible region, whereas these fractions decrease to $\sim40\%$ ($\sim50\%$, $\sim70\%$) for \seob{} recovery. Consistently, the fraction of events for which the most probable host coincides with the true host is $\sim40\%$ for \imrxpnr{} recovery, compared to $\sim20\%$ for \seob{}.

Taken together, these results demonstrate that
\begin{inparaenum}[(1)]
    \item waveform-induced localization biases can substantially impact host-galaxy association, even when the overall sky areas are comparable; and
    \item the tendency of the posterior inference process to compensate waveform inaccuracies through parameter adjustment can lead to biased inferences, consistent with our ansätze in Sec.~\ref{sec:wsys}.
\end{inparaenum}
We emphasize, however, that an incorrect host identification does not necessarily imply a biased \(H_0\) estimate even at the single-event level as we show next.

%%%%%%%%%%%%%%%%%%%%%%%%%%%%%%%%%%%%%%%%%%%%%%%%%%%%%%%%%%%%%%%%%%%%%%%%
    \section{Observational Imperatives for Hubble constant measurements}
%%%%%%%%%%%%%%%%%%%%%%%%%%%%%%%%%%%%%%%%%%%%%%%%%%%%%%%%%%%%%%%%%%%%%%%%
\begin{figure*}
    \centering
    \includegraphics[width=\linewidth]{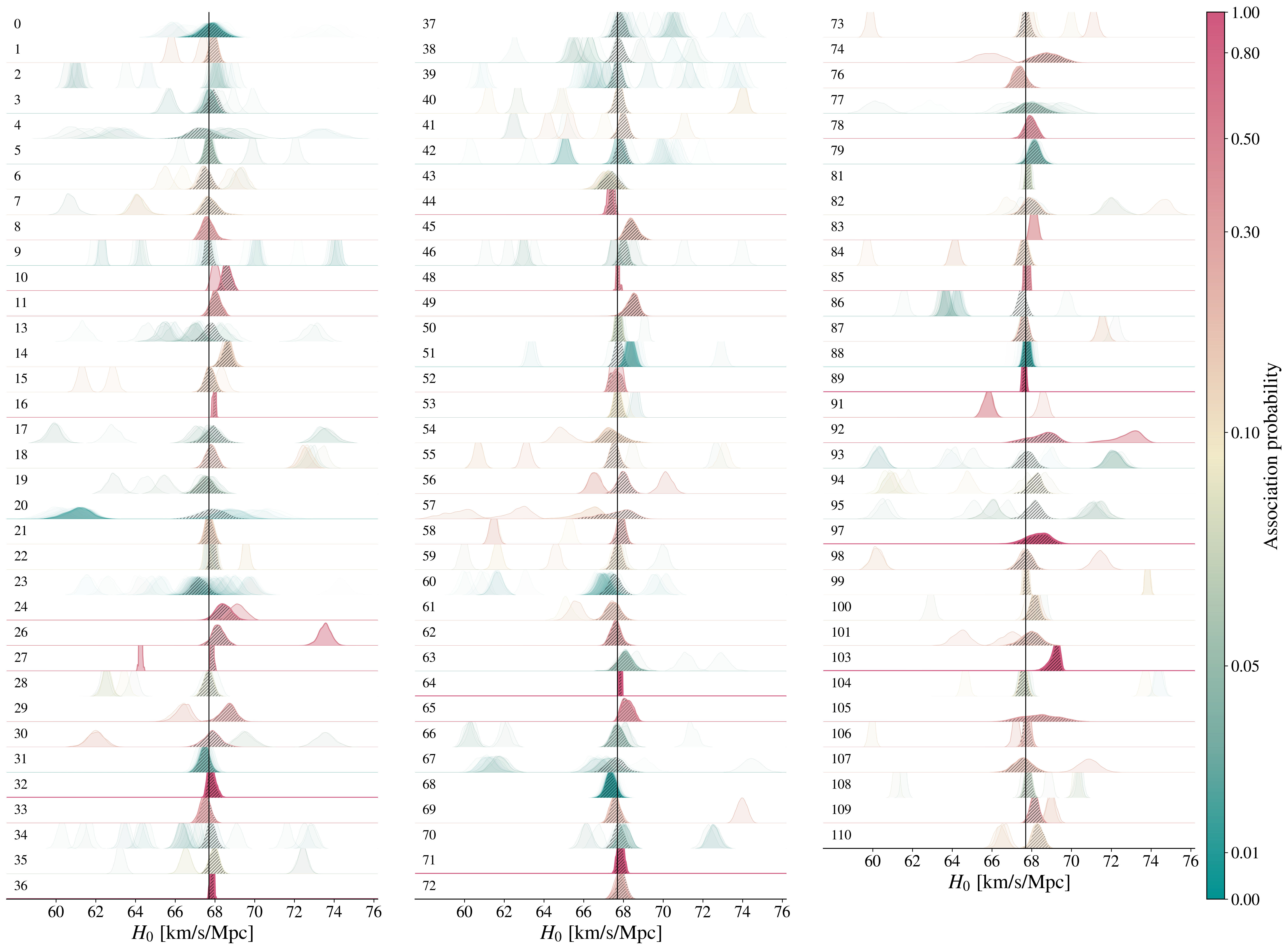}
    \caption{$H_0$ posterior distribution for recovery with \imrxpnr{}. Association probability shown with colorscale and opacity. True (simulated) counterpart hatched. Injected $H_0=67.7 ~\text{km}/\text{s}/\text{Mpc}$ marked with the vertical line.} 
    \label{fig:hubble-ridge}
\end{figure*}

\begin{figure}
    \centering
    \includegraphics[width=0.95\linewidth]{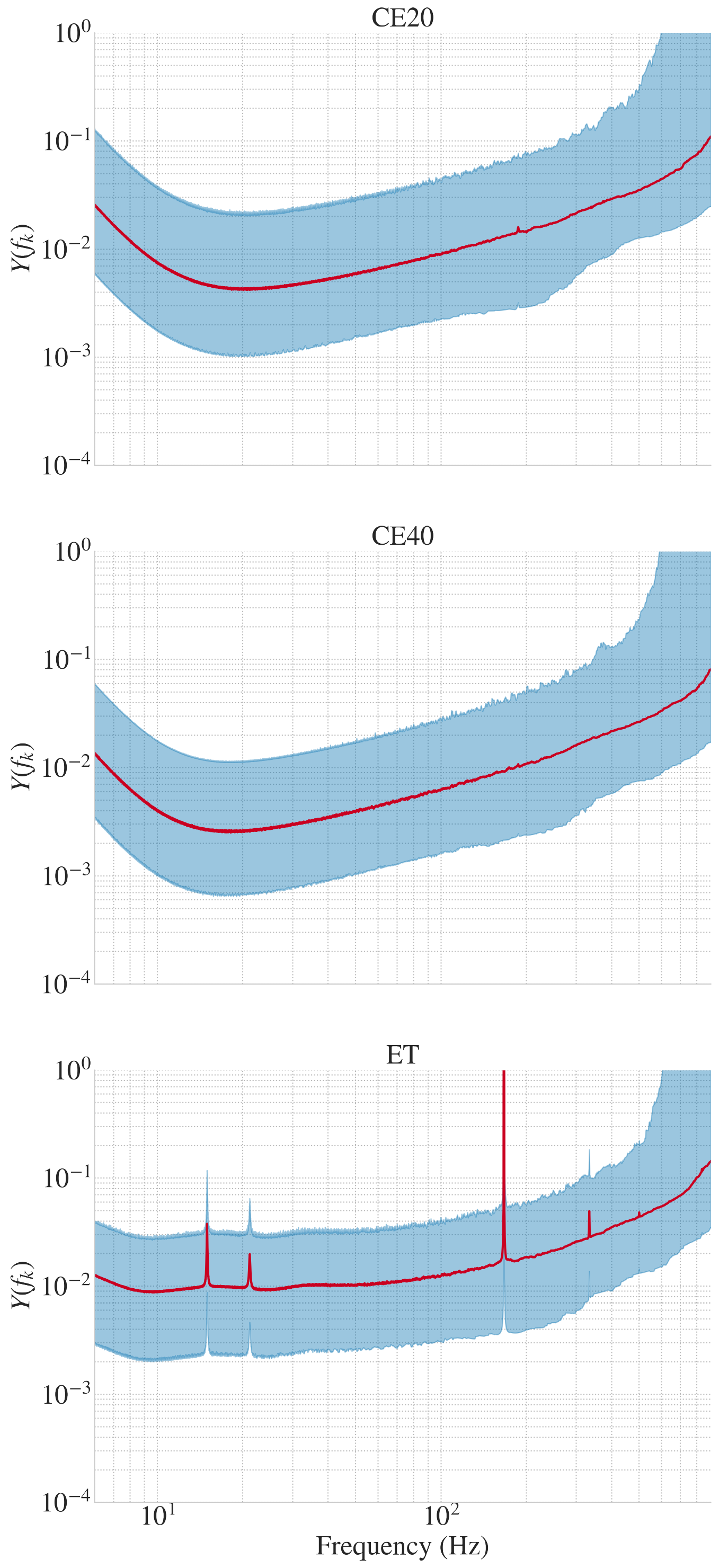}
    \caption{Characteristic frequency-dependent calibration tolerance \(Y(f_k)\) for representative golden dark siren binary population observed with \ac{ce}40, \ac{ce}20 and \ac{et} (harmonic-mean). The solid red curve shows the median behavior while the shaded region indicates the 90\% spread across the signal population. For comparison, the current Advanced LIGO detectors typically have both amplitude and phase calibration uncertainty at \(\sim 10^{-2}\) level~\citep{Capote:2024rmo}.}
    \label{fig:calibration}
\end{figure}
The localization biases discussed above are observationally relevant only to the extent that they propagate into biased \(H_0\) measurement. We therefore assess their impact on single-event measurements of the Hubble constant by combining the \ac{gw} posteriors with candidate \ac{em} host galaxies under the same cosmological and population assumptions introduced in the previous section. The redshifts are assumed to be known without uncertainty; this is a reasonable hypothesis, since a spectroscopical measurement of redshift with typical error of $\approx 50$ km/s at 600-800 Mpc (where the majority of the most informative \gds{} are located) yields to $\lesssim 0.1\%$ uncertainty on $z$. We verify that the corresponding variations in association probabilities are negligible. In addition, this effect is subdominant compared to the proper motion of the galaxies, which we choose to ignore to isolate biases exclusively caused by waveform inaccuracies.

Figure~\ref{fig:hubble-ridge} shows the resulting \(H_0\) posteriors for \imrxpnr{} signal recovery (see Appendix~\ref{sec:hubble} for \seob{}). Each curve corresponds to a potential host galaxy and is weighted by its association probability inferred from the three-dimensional localization (Appendix~\ref{sec:3d}). The color scale and opacity encode these probabilities, while the hatched curve denotes the posterior obtained when the true host galaxy is assumed to be known. In most cases, the recovered $H_0$ posteriors contain the true value, with widths that scale inversely with the event \ac{snr}, indicating that waveform-induced localization biases do not generically translate into biased \(H_0\) inference.

Importantly, this conclusion holds even for events that do not qualify as \gds{} due to multiple viable host associations. Although waveform systematics can broaden the set of plausible hosts or redistribute probability away from the true galaxy, the redshift of the most probable host encloses the true $H_0$ within the $67\%$ credible interval in $60\%$ of cases for \nrhyb{}--\imrst{} recovery and $42\%$ for \nrhyb{}--\seob{}. Moreover, in many systems (e.g., sources~0, 68, and~88 in Fig.~\ref{fig:hubble-ridge}), several low-probability associations belong to the same galaxy group or cluster. In these cases, the corresponding \(H_0\) posteriors are nearly identical and collectively sum to unit probability, rendering the event nearly as informative as a true \gds{} (cf.\ sources~32, 36 and~71).

A quantitative summary is provided in Table~\ref{tab:reconstruction_statistic}. For \imrxpnr{} (\seob{}) signal reconstruction, out of $103$ ($93$) systems, $40\%$ ($18\%$) are correctly associated with the true host galaxy. Of these, only $17\%$ ($24\%$) are strict \gds{} ($p_\text{association}=1$, see App. \ref{sec:3d}), while $25\%$ ($12\%$) qualify as \textit{copper sirens} ($p_\text{association}>0.5$). Among correctly identified hosts, $70\%$ ($65\%$) enclose the true $H_0$ within the $90\%$ credible interval, which we take as a criterion for successful $H_0$ reconstruction. For the remaining $30\%$ ($35\%$), biases in the luminosity distance are sufficient to shift the inferred $H_0$ away from the true value.

Conversely, $53\%$ ($37\%$) of misidentified hosts nonetheless yield correct $H_0$ reconstruction, illustrating that incorrect galaxy association does not necessarily imply biased cosmological inference. Overall, the most probable host leads to correct $H_0$ reconstruction in $60\%$ ($42\%)$ of cases. For \textit{standard dark sirens}, where multiple low-probability counterparts collectively dominate the association probability, we find that $53\%$ ($31\%)$ of systems yield correct $H_0$ reconstruction once the cumulative probability exceeds $50\%$. Finally, while ``impostor \gds{}''---systems with a single incorrect counterpart---are absent in the \imrxpnr{} analysis, they occur in the \seob{} recovery at a rate comparable to that of genuine \gds.

\begin{table}
\centering
\begin{tabular}{lccc}
\hline\hline
Model & Correct $H_0$ & Incorrect $H_0$ & Total \\
\hline
\multicolumn{4}{c}{Correct host association} \\
\imrxpnr{} & 29 (28\%) & 12 (12\%) & 41 (40\%) \\
\seob{}   & 11 (12\%) &  6 (6\%)  & 17 (18\%) \\
\hline
\multicolumn{4}{c}{Incorrect host association} \\
\imrxpnr{} & 33 (32\%) & 29 (28\%) & 62 (60\%) \\
\seob{}   & 28 (30\%) & 48 (52\%) & 76 (82\%) \\
\hline
\multicolumn{4}{c}{Total} \\
\imrxpnr{} & 62 (60\%) & 41 (40\%) & 103 (100\%) \\
\seob{}   & 39 (42\%) & 54 (58\%) & 93 (100\%) \\
\hline\hline
\end{tabular}
\caption{
Host association and $H_0$ reconstruction statistics for the two waveform models.
Percentages are relative to the total number of events for each model.
}
\label{tab:reconstruction_statistic}
\end{table}

From a data‑analysis perspective, these results highlight a key observational imperative: accurate cosmological inference does not require unambiguous host identification, but rather a self‑consistent mapping between luminosity distance and the redshift distribution of plausible hosts. Provided that flat $\Lambda$CDM is a valid description of the Universe and that current quasi‑circular waveform models adequately represent high‑\ac{snr} signals, waveform‑induced host‑association ambiguities do not, by themselves, preclude robust $H_0$ measurements with next-generation \ac{gw} detectors.

We do not perform a population-level cosmological inference in this work. Our Bayesian analysis is carried out in the absence of noise to isolate the impact of waveform systematics on luminosity–distance inference and host association. In this regime, the likelihoods do not represent independent realizations and therefore cannot be consistently combined within a hierarchical framework. Moreover, all systems considered lie well within the sensitive volume of the detector network, such that the detection probability is effectively unity and selection effects are nearly constant across the relevant parameter space.

%%%%%%%%%%%%%%%%%%%%%%%%%%%%%%%%%%%%%%%%%
\section{How good should the detector calibrations be?}
%%%%%%%%%%%%%%%%%%%%%%%%%%%%%%%%%%%%%%%%%

Throughout this work, we have implicitly assumed that the detector data are known exactly. In practice, the calibrated strain is affected by frequency‑dependent amplitude and phase uncertainties arising from imperfect knowledge of the detector response. While calibration systematics are not the focus of this study, they can, in principle, bias host‑galaxy association and cosmological inference for high‑\ac{snr} events observed with \ac{xg} detectors. So, we give a back-of-the-envelope estimate of the level at which calibration uncertainties remain subdominant to statistical fluctuations. We follow an indistinguishability argument: calibration errors should not introduce discrepancies large enough to be distinguishable in the likelihood. Under idealized conditions—perfect waveform modeling, known source parameters, and noise‑free data—this requirement leads to a simple bound on the combined fractional amplitude and phase error, weighted by the signal‑to‑noise ratio (see Appendix~\ref{app:calibration} for derivation).

The resulting condition may be expressed as:
\begin{equation}
    \sum_k \rho_k^2\,
    \bigl[\delta A^2(f_k)+\delta\phi^2(f_k)\bigr] < 1 ,
\label{eq:calib-global}
\end{equation}
where $\rho_k^2$ denotes per frequency \ac{snr} contribution and  $\delta A$ and $\delta\phi$ denote fractional amplitude and phase offsets at the same frequency. Equation~\eqref{eq:calib-global} shows that calibration accuracy must be most stringent in the frequency regions that dominate the \ac{snr}. A sufficient but overly aggressive, frequency‑resolved sufficient condition is
\begin{equation}\label{eq:calib-local}
    Y(f_k)^2 = \delta A^2(f_k)+\delta\phi^2(f_k) \lesssim \rho_k^{-2}/N,
\end{equation}
which defines an envelope for the maximum tolerable calibration error as a function of frequency.

Fig.~\ref{fig:calibration} shows this tolerance for representative golden dark siren signals in \ac{xg} detectors. For high‑\ac{snr} events, the most stringent requirements occur at low frequencies, where the signal accumulates most of its \ac{snr}. In selected bands of \ac{ce}40, the required accuracy can reach $\sim 5 \times 10^{-4} $, significantly more stringent than the $\mathcal{O}(10^{-2})$ calibration uncertainties typical of current detectors. This highlights the need for substantially improved calibration accuracy to exploit \gds{} cosmology with next‑generation observatories fully.

Although derived here for calibration systematics, this criterion applies more generally to any measurement error that can be expressed as fractional amplitude and phase deviations, and therefore provides a conservative benchmark for waveform‑model accuracy as well.

%%%%%%%%%%%%%%%%%%%%%%%%%%%%%%%%%%%%%%%%%%%%%%%%%%%%%
\section{Conclusions and Future Prospects}
%%%%%%%%%%%%%%%%%%%%%%%%%%%%%%%%%%%%%%%%%%%%%%%%%%%%%

Golden dark sirens observed with next-generation (XG) gravitational-wave detectors promise an unprecedented opportunity for precision cosmology, potentially enabling sub-percent measurements of the Hubble constant without reliance on electromagnetic counterparts. In this work, we have examined a key limiting factor for realizing this potential: systematic errors arising from waveform modeling and detector calibration in the high--signal-to-noise-ratio regime characteristic of XG observations.

Using a synthetic population consistent with \citet{KAGRA:2021duu} of nearby binary black holes and numerical-relativity–based reference waveforms, we have quantified waveform inaccuracies from both modeling and data-analysis perspectives and traced their impact through single-event parameter estimation, three-dimensional source localization, host-galaxy association, and ultimately $H_0$ inference. Our results demonstrate that, while current state-of-the-art waveform models are often capable of yielding statistically consistent $H_0$ posteriors, waveform-induced biases at high SNR can significantly distort sky localization and luminosity-distance estimates. These distortions, in turn, affect host-galaxy ranking and can lead to incorrect redshift assignments for a non-negligible fraction of events.

An important and reassuring outcome of this study is that imperfect host identification does not, in general, imply biased cosmological inference. In many cases, waveform systematics broaden or reshuffle the set of plausible host galaxies without driving the inferred $H_0$ away from its true value. This is particularly true when multiple candidate hosts belong to the same galaxy group or cluster, so that their redshifts are effectively interchangeable for low-redshift cosmology. In this sense, golden dark siren cosmology is more robust to modest localization biases than might be inferred from host-identification metrics alone. What matters observationally is not unique host identification per se, but a self-consistent mapping between the GW-inferred luminosity distance and the redshift distribution of plausible hosts.

At the same time, our analysis makes clear that the statistical power of golden dark sirens places stringent demands on controlling systematic errors. Waveform inaccuracies must scale at least as $\mathcal{O}(\rho^{-2})$ with signal-to-noise ratio to remain subdominant to statistical uncertainties, and similar considerations apply to detector calibration. Our order-of-magnitude estimates indicate that calibration accuracy of $10^{-4}$ in the most informative frequency bands may be required for the loudest XG events, far exceeding the standards set by current detectors. These requirements underscore the need for sustained advances in waveform modeling, numerical relativity, and calibration strategies if the full cosmological potential of XG detectors is to be realized.

Several important questions raised by this work remain to be addressed. In particular, our host-association analysis relies on mock galaxy catalogs and idealized assumptions about catalog completeness. A natural and necessary next step is to repeat this study using real, deep galaxy surveys with well-characterized selection functions. Galaxy fields such as \textsc{SHELA} and \textsc{COSMOS}, as observed by HETDEX \citep{Dang:2025vqx}, provide an ideal testbed for this purpose. By constructing a galaxy catalog complete to limiting $K$-band magnitudes of $K \simeq 23$--24, one can realistically assess how catalog incompleteness and misidentification of host galaxies bias golden dark siren cosmology in practice.

Such an analysis will allow us to quantify, in a fully data-driven manner, the extent to which waveform-induced biases in three-dimensional localization propagate into systematic errors on $H_0$, and to test explicitly the hypothesis suggested by this work: that small localization biases are largely benign when the inferred host galaxy lies within the same group or cluster as the true host. We will undertake this investigation in a forthcoming paper to establish robust, end-to-end requirements on waveform accuracy, calibration, and galaxy-catalog depth for precision cosmology with golden dark sirens in the XG era.

%%%%%%%%%%%%%%%%%%%%%%%%%%%%%%%%%%%%%%%%%%%%%%%%%%%%%
\section{Acknowledgments}
%%%%%%%%%%%%%%%%%%%%%%%%%%%%%%%%%%%%%%%%%%%%%%%%%%%%%
KC acknowledges the use of OpenAI’s GPT‑5.2 for assistance with improving the presentation of figures and thanks Ingrid Helene Håvik for musical motivation during late‑night debugging. 
He also thanks Rossella Gamba for one-way conversations about this work and for listening to more discussions about it than anyone reasonably should. The authors thank Arnab Dhani and Jonathan Gair for their valuable comments and suggestions. We also thank Charlie Hoy for sharing the script to calculate the sky area. The authors acknowledge support from NSF grants PHY-2207638, AST-2307147, PHY-2308886, and PHY-2309064. GB acknowledges support from INFN grant 27626 (INFN-NSF/LIGO exchange fellowship). The authors acknowledge the use of the Gwave (PSU) and LDAS (CIT) clusters for computational/numerical work. 

\bibliography{references}

@article{Gupta:2023lga,
    author = "Gupta, Ish and others",
    title = "{Characterizing gravitational wave detector networks: from A$^\sharp$ to cosmic explorer}",
    eprint = "2307.10421",
    archivePrefix = "arXiv",
    primaryClass = "gr-qc",
    reportNumber = "CE Document No. P2300019, CE Document No. P2300019-v2",
    doi = "10.1088/1361-6382/ad7b99",
    journal = "Class. Quant. Grav.",
    volume = "41",
    number = "24",
    pages = "245001",
    year = "2024"
}

@article{VIRGO:2014yos,
    author = "Acernese, F. and others",
    collaboration = "VIRGO",
    title = "{Advanced Virgo: a second-generation interferometric gravitational wave detector}",
    eprint = "1408.3978",
    archivePrefix = "arXiv",
    primaryClass = "gr-qc",
    doi = "10.1088/0264-9381/32/2/024001",
    journal = "Class. Quant. Grav.",
    volume = "32",
    number = "2",
    pages = "024001",
    year = "2015"
}

@article{LIGOScientific:2014pky,
    author = "Aasi, J. and others",
    collaboration = "{LIGO Scientific}",
    title = "{Advanced LIGO}",
    eprint = "1411.4547",
    archivePrefix = "arXiv",
    primaryClass = "gr-qc",
    doi = "10.1088/0264-9381/32/7/074001",
    journal = "Class. Quant. Grav.",
    volume = "32",
    pages = "074001",
    year = "2015"
}

@article{LIGOScientific:2017vwq,
    author = "Abbott, B. P. and others",
    collaboration = "{LIGO Scientific Collaboration and Virgo Collaboration}",
    title = "{GW170817: Observation of Gravitational Waves from a Binary Neutron Star Inspiral}",
    eprint = "1710.05832",
    archivePrefix = "arXiv",
    primaryClass = "gr-qc",
    reportNumber = "LIGO-P170817",
    doi = "10.1103/PhysRevLett.119.161101",
    journal = "Phys. Rev. Lett.",
    volume = "119",
    number = "16",
    pages = "161101",
    year = "2017"
}

@article{LIGOScientific:2017adf,
    author = "Abbott, B. P. and others",
    collaboration = "LIGO Scientific, Virgo, 1M2H, Dark Energy Camera GW-E, DES, DLT40, Las Cumbres Observatory, VINROUGE, MASTER",
    title = "{A gravitational-wave standard siren measurement of the Hubble constant}",
    eprint = "1710.05835",
    archivePrefix = "arXiv",
    primaryClass = "astro-ph.CO",
    reportNumber = "LIGO-P1700296, FERMILAB-PUB-17-472-A-AE",
    doi = "10.1038/nature24471",
    journal = "Nature",
    volume = "551",
    number = "7678",
    pages = "85--88",
    year = "2017"
}

@unpublished{Dang:2025vqx,
    author = "Dang, Yixuan and others",
    title = "{Golden and Silver Dark Sirens for precise H0 measurement with HETDEX}",
    eprint = "2512.21729",
    archivePrefix = "arXiv",
    primaryClass = "astro-ph.CO",
    month = "12",
    year = "2025",
    note = {\url{https://arxiv.org/pdf/2512.21729}}
}

@article{Schutz:1986gp,
    author = "Schutz, Bernard F.",
    title = "{Determining the Hubble Constant from Gravitational Wave Observations}",
    doi = "10.1038/323310a0",
    journal = "Nature",
    volume = "323",
    pages = "310--311",
    year = "1986"
}

@article{Borhanian:2020vyr,
    author = "Borhanian, Ssohrab and Dhani, Arnab and Gupta, Anuradha and Arun, K. G. and Sathyaprakash, B. S.",
    title = "{Dark Sirens to Resolve the Hubble{\textendash}Lema{\^\i}tre Tension}",
    eprint = "2007.02883",
    archivePrefix = "arXiv",
    primaryClass = "astro-ph.CO",
    reportNumber = "LIGO document number LIGO-P2000229",
    doi = "10.3847/2041-8213/abcaf5",
    journal = "Astrophys. J. Lett.",
    volume = "905",
    number = "2",
    pages = "L28",
    year = "2020"
}

@article{Leandro:2021qlc,
    author = "Leandro, Hebertt and Marra, Valerio and Sturani, Riccardo",
    title = "{Measuring the Hubble constant with black sirens}",
    eprint = "2109.07537",
    archivePrefix = "arXiv",
    primaryClass = "gr-qc",
    doi = "10.1103/PhysRevD.105.023523",
    journal = "Phys. Rev. D",
    volume = "105",
    number = "2",
    pages = "023523",
    year = "2022"
}

@article{Gair:2022zsa,
    author = "Gair, Jonathan R. and others",
    title = "{The Hitchhiker{\textquoteright}s Guide to the Galaxy Catalog Approach for Dark Siren Gravitational-wave Cosmology}",
    eprint = "2212.08694",
    archivePrefix = "arXiv",
    primaryClass = "gr-qc",
    doi = "10.3847/1538-3881/acca78",
    journal = "Astron. J.",
    volume = "166",
    number = "1",
    pages = "22",
    year = "2023"
}

@article{Chandra:2024dhf,
    author = "Chandra, Koustav",
    title = "{gwforge: a user-friendly package to generate gravitational-wave mock data}",
    eprint = "2407.21109",
    archivePrefix = "arXiv",
    primaryClass = "gr-qc",
    doi = "10.1088/1361-6382/ad9b68",
    journal = "Class. Quant. Grav.",
    volume = "42",
    number = "2",
    pages = "025003",
    year = "2025"
}

@article{KAGRA:2021duu,
    author = "Abbott, R. and others",
    collaboration = "KAGRA, VIRGO, LIGO Scientific",
    title = "{Population of Merging Compact Binaries Inferred Using Gravitational Waves through GWTC-3}",
    eprint = "2111.03634",
    archivePrefix = "arXiv",
    primaryClass = "astro-ph.HE",
    reportNumber = "LIGO-P2100239 ; Data release: https://zenodo.org/record/5655785, LIGO-P2100239",
    doi = "10.1103/PhysRevX.13.011048",
    journal = "Phys. Rev. X",
    volume = "13",
    number = "1",
    pages = "011048",
    year = "2023"
}

@article{Dupletsa:2022scg,
    author = "Dupletsa, Ulyana and Harms, Jan and Banerjee, Biswajit and Branchesi, Marica and Goncharov, Boris and Maselli, Andrea and Oliveira, Ana Carolina Silva and Ronchini, Samuele and Tissino, Jacopo",
    title = "{gwfish: A simulation software to evaluate parameter-estimation capabilities of gravitational-wave detector networks}",
    eprint = "2205.02499",
    archivePrefix = "arXiv",
    primaryClass = "gr-qc",
    doi = "10.1016/j.ascom.2022.100671",
    journal = "Astron. Comput.",
    volume = "42",
    pages = "100671",
    year = "2023"
}

@article{Fosalba:2013wxa,
    author = "Fosalba, P. and Crocce, M. and Gazta{\~n}aga, E. and Castander, F. J.",
    title = "{The MICE grand challenge lightcone simulation {\textendash} I. Dark matter clustering}",
    eprint = "1312.1707",
    archivePrefix = "arXiv",
    primaryClass = "astro-ph.CO",
    doi = "10.1093/mnras/stv138",
    journal = "Mon. Not. Roy. Astron. Soc.",
    volume = "448",
    number = "4",
    pages = "2987--3000",
    year = "2015"
}

@article{Crocce:2013vda,
    author = "Crocce, M. and Castander, F. J. and Gaztanaga, E. and Fosalba, P. and Carretero, J.",
    title = "{The MICE Grand Challenge lightcone simulation {\textendash} II. Halo and galaxy catalogues}",
    eprint = "1312.2013",
    archivePrefix = "arXiv",
    primaryClass = "astro-ph.CO",
    doi = "10.1093/mnras/stv1708",
    journal = "Mon. Not. Roy. Astron. Soc.",
    volume = "453",
    number = "2",
    pages = "1513--1530",
    year = "2015"
}

@article{Fosalba:2013mra,
    author = "Fosalba, P. and Gazta{\~n}aga, E. and Castander, F. J. and Crocce, M.",
    title = "{The MICE Grand Challenge light-cone simulation {\textendash} III. Galaxy lensing mocks from all-sky lensing maps}",
    eprint = "1312.2947",
    archivePrefix = "arXiv",
    primaryClass = "astro-ph.CO",
    doi = "10.1093/mnras/stu2464",
    journal = "Mon. Not. Roy. Astron. Soc.",
    volume = "447",
    number = "2",
    pages = "1319--1332",
    year = "2015"
}

@article{Carretero:2014ltj,
    author = "Carretero, J. and Castander, F. J. and Gaztanaga, E. and Crocce, M. and Fosalba, P.",
    title = "{An algorithm to build mock galaxy catalogues using MICE simulations}",
    eprint = "1411.3286",
    archivePrefix = "arXiv",
    primaryClass = "astro-ph.GA",
    doi = "10.1093/mnras/stu2402",
    journal = "Mon. Not. Roy. Astron. Soc.",
    volume = "447",
    pages = "650",
    year = "2015"
}

@article{Hoffmann:2014ida,
    author = "Hoffmann, Kai and Bel, Julien and Gazta{\~n}aga, Enrique and Crocce, Martin and Fosalba, Pablo and Castander, Francisco J.",
    title = "{Measuring the growth of matter fluctuations with third-order galaxy correlations}",
    eprint = "1403.1259",
    archivePrefix = "arXiv",
    primaryClass = "astro-ph.CO",
    doi = "10.1093/mnras/stu2492",
    journal = "Mon. Not. Roy. Astron. Soc.",
    volume = "447",
    number = "2",
    pages = "1724--1745",
    year = "2015"
}

@article{Dhani:2024jja,
    author = {Dhani, Arnab and V{\"o}lkel, Sebastian and Buonanno, Alessandra and Estelles, Hector and Gair, Jonathan and Pfeiffer, Harald P. and Pompili, Lorenzo and Toubiana, Alexandre},
    title = "{Systematic Biases in Estimating the Properties of Black Holes Due to Inaccurate Gravitational-Wave Models}",
    eprint = "2404.05811",
    archivePrefix = "arXiv",
    primaryClass = "gr-qc",
    month = "4",
    year = "2024"
}

@article{Cutler:2007mi,
    author = "Cutler, Curt and Vallisneri, Michele",
    title = "{LISA detections of massive black hole inspirals: Parameter extraction errors due to inaccurate template waveforms}",
    eprint = "0707.2982",
    archivePrefix = "arXiv",
    primaryClass = "gr-qc",
    doi = "10.1103/PhysRevD.76.104018",
    journal = "Phys. Rev. D",
    volume = "76",
    pages = "104018",
    year = "2007"
}

@article{Dhani:2025xgt,
    author = "Dhani, Arnab and Gair, Jonathan and Buonanno, Alessandra",
    title = "{The fault in our sirens: Hierarchical diagnosis of waveform systematics in Hubble-Lema{\^\i}tre constant measurements}",
    eprint = "2507.11278",
    archivePrefix = "arXiv",
    primaryClass = "gr-qc",
    month = "7",
    year = "2025"
}

@article{Toubiana:2024car,
    author = "Toubiana, Alexandre and Gair, Jonathan R.",
    title = "{Indistinguishability criterion and estimating the presence of biases}",
    eprint = "2401.06845",
    archivePrefix = "arXiv",
    primaryClass = "gr-qc",
    month = "1",
    year = "2024"
}

@article{Purrer:2019jcp,
    author = {P{\"u}rrer, Michael and Haster, Carl-Johan},
    title = "{Gravitational waveform accuracy requirements for future ground-based detectors}",
    eprint = "1912.10055",
    archivePrefix = "arXiv",
    primaryClass = "gr-qc",
    doi = "10.1103/PhysRevResearch.2.023151",
    journal = "Phys. Rev. Res.",
    volume = "2",
    number = "2",
    pages = "023151",
    year = "2020"
}

@article{Chen:2024gdn,
    author = "Chen, Hsin-Yu and Ezquiaga, Jose Mar\'\i{}a and Gupta, Ish",
    title = "{Cosmography with next-generation gravitational wave detectors}",
    eprint = "2402.03120",
    archivePrefix = "arXiv",
    primaryClass = "gr-qc",
    doi = "10.1088/1361-6382/ad424f",
    journal = "Class. Quant. Grav.",
    volume = "41",
    number = "12",
    pages = "125004",
    year = "2024"
}

@article{Vitale:2011wu,
    author = "Vitale, Salvatore and Del Pozzo, Walter and Li, Tjonnie G. F. and Van Den Broeck, Chris and Mandel, Ilya and Aylott, Ben and Veitch, John",
    title = "{Effect of calibration errors on Bayesian parameter estimation for gravitational wave signals from inspiral binary systems in the Advanced Detectors era}",
    eprint = "1111.3044",
    archivePrefix = "arXiv",
    primaryClass = "gr-qc",
    doi = "10.1103/PhysRevD.85.064034",
    journal = "Phys. Rev. D",
    volume = "85",
    pages = "064034",
    year = "2012"
}

@article{Essick:2022vzl,
    author = "Essick, Reed",
    title = "{Calibration uncertainty{\textquoteright}s impact on gravitational-wave observations}",
    eprint = "2202.00823",
    archivePrefix = "arXiv",
    primaryClass = "astro-ph.IM",
    doi = "10.1103/PhysRevD.105.082002",
    journal = "Phys. Rev. D",
    volume = "105",
    number = "8",
    pages = "082002",
    year = "2022"
}

@article{Capote:2024mqe,
    author = "Capote, Elenna and Dartez, Louis and Davis, Derek",
    title = "{Technical noise, data quality, and calibration requirements for next-generation gravitational-wave science}",
    eprint = "2404.04761",
    archivePrefix = "arXiv",
    primaryClass = "astro-ph.IM",
    reportNumber = "CE-P2400001",
    doi = "10.1088/1361-6382/ad694d",
    journal = "Class. Quant. Grav.",
    volume = "41",
    number = "18",
    pages = "185001",
    year = "2024"
}

@article{Hanselman:2024hqy,
    author = "Hanselman, Alexandra G. and Vijaykumar, Aditya and Fishbach, Maya and Holz, Daniel E.",
    title = "{Gravitational-wave Dark Siren Cosmology Systematics from Galaxy Weighting}",
    eprint = "2405.14818",
    archivePrefix = "arXiv",
    primaryClass = "astro-ph.CO",
    doi = "10.3847/1538-4357/ad9393",
    journal = "Astrophys. J.",
    volume = "979",
    number = "1",
    pages = "9",
    year = "2025"
}

@article{Harry:2017weg,
    author = "Harry, Ian and Calder{\'o}n Bustillo, Juan and Nitz, Alex",
    title = "{Searching for the full symphony of black hole binary mergers}",
    eprint = "1709.09181",
    archivePrefix = "arXiv",
    primaryClass = "gr-qc",
    reportNumber = "LIGO-DOCUMENT-P1700262, LIGO Document P1700262",
    doi = "10.1103/PhysRevD.97.023004",
    journal = "Phys. Rev. D",
    volume = "97",
    number = "2",
    pages = "023004",
    year = "2018"
}

@article{Chandra:2022ixv,
    author = "Chandra, Koustav and Calder{\'o}n Bustillo, Juan and Pai, Archana and Harry, I. W.",
    title = "{First gravitational-wave search for intermediate-mass black hole mergers with higher-order harmonics}",
    eprint = "2207.01654",
    archivePrefix = "arXiv",
    primaryClass = "gr-qc",
    reportNumber = "LIGO-P2200182",
    doi = "10.1103/PhysRevD.106.123003",
    journal = "Phys. Rev. D",
    volume = "106",
    number = "12",
    pages = "123003",
    year = "2022"
}

@article{Buonanno:1998gg,
    author = "Buonanno, A. and Damour, T.",
    title = "{Effective one-body approach to general relativistic two-body dynamics}",
    eprint = "gr-qc/9811091",
    archivePrefix = "arXiv",
    reportNumber = "IHES-P-98-74",
    doi = "10.1103/PhysRevD.59.084006",
    journal = "Phys. Rev. D",
    volume = "59",
    pages = "084006",
    year = "1999"
}

@article{Buonanno:2000ef,
    author = "Buonanno, Alessandra and Damour, Thibault",
    title = "{Transition from inspiral to plunge in binary black hole coalescences}",
    eprint = "gr-qc/0001013",
    archivePrefix = "arXiv",
    reportNumber = "IHES-P-99-90, GRP-99-521",
    doi = "10.1103/PhysRevD.62.064015",
    journal = "Phys. Rev. D",
    volume = "62",
    pages = "064015",
    year = "2000"
}

@article{Read:2023hkv,
    author = "Read, Jocelyn S.",
    title = "{Waveform uncertainty quantification and interpretation for gravitational-wave astronomy}",
    eprint = "2301.06630",
    archivePrefix = "arXiv",
    primaryClass = "gr-qc",
    doi = "10.1088/1361-6382/acd29b",
    journal = "Class. Quant. Grav.",
    volume = "40",
    number = "13",
    pages = "135002",
    year = "2023"
}

@article{LIGOScientific:2016xax,
    author = "Abbott, B. P. and others",
    collaboration = "LIGO Scientific",
    title = "{Calibration of the Advanced LIGO detectors for the discovery of the binary black-hole merger GW150914}",
    eprint = "1602.03845",
    archivePrefix = "arXiv",
    primaryClass = "gr-qc",
    doi = "10.1103/PhysRevD.95.062003",
    journal = "Phys. Rev. D",
    volume = "95",
    number = "6",
    pages = "062003",
    year = "2017"
}

@article{LIGOScientific:2017aaj,
    author = "Cahillane, Craig and others",
    collaboration = "LIGO Scientific",
    title = "{Calibration uncertainty for Advanced LIGO{\textquoteright}s first and second observing runs}",
    eprint = "1708.03023",
    archivePrefix = "arXiv",
    primaryClass = "astro-ph.IM",
    doi = "10.1103/PhysRevD.96.102001",
    journal = "Phys. Rev. D",
    volume = "96",
    number = "10",
    pages = "102001",
    year = "2017"
}

@article{Sun:2021qcg,
    author = "Sun, Ling and others",
    title = "{Characterization of systematic error in Advanced LIGO calibration in the second half of O3}",
    eprint = "2107.00129",
    archivePrefix = "arXiv",
    primaryClass = "astro-ph.IM",
    month = "6",
    year = "2021"
}

@article{Capote:2024rmo,
    author = "Capote, E. and others",
    title = "{Advanced LIGO detector performance in the fourth observing run}",
    eprint = "2411.14607",
    archivePrefix = "arXiv",
    primaryClass = "gr-qc",
    reportNumber = "LIGO-P2400256",
    doi = "10.1103/PhysRevD.111.062002",
    journal = "Phys. Rev. D",
    volume = "111",
    number = "6",
    pages = "062002",
    year = "2025"
}

@techreport{Farr2014Calibration,
  author       = {Farr, Will M. and Farr, Benjamin and Littenberg, Tyson},
  title        = {Calibration Errors in {CBC} Waveforms},
  institution  = {LIGO Laboratory},
  number       = {T1400682},
  year         = {2014},
  month        = oct,
  note         = {Draft version, October 22, 2014},
  url          = {https://dcc.ligo.org/public/0116/T1400682/001/calnote.pdf}
}

@article{Huang:2022rdg,
    author = "Huang, Yiwen and Chen, Hsin-Yu and Haster, Carl-Johan and Sun, Ling and Vitale, Salvatore and Kissel, Jeffrey S.",
    title = "{Impact of calibration uncertainties on Hubble constant measurements from gravitational-wave sources}",
    eprint = "2204.03614",
    archivePrefix = "arXiv",
    primaryClass = "gr-qc",
    doi = "10.1103/PhysRevD.111.063034",
    journal = "Phys. Rev. D",
    volume = "111",
    number = "6",
    pages = "063034",
    year = "2025"
}

@article{Holz:2005df,
    author = "Holz, Daniel E. and Hughes, Scott A.",
    title = "{Using gravitational-wave standard sirens}",
    eprint = "astro-ph/0504616",
    archivePrefix = "arXiv",
    doi = "10.1086/431341",
    journal = "Astrophys. J.",
    volume = "629",
    pages = "15--22",
    year = "2005"
}

@article{LIGOScientific:2025pvj,
    author = "Abac, A. G. and others",
    collaboration = "LIGO Scientific, VIRGO, KAGRA",
    title = "{GWTC-4.0: Population Properties of Merging Compact Binaries}",
    eprint = "2508.18083",
    archivePrefix = "arXiv",
    primaryClass = "astro-ph.HE",
    reportNumber = "LIGO-P2400004",
    month = "8",
    year = "2025"
}

@article{Ashton:2020kyr,
    author = "Ashton, Gregory and Ackley, Kendall and Hernandez, Ignacio Maga{\~n}a and Piotrzkowski, Brandon",
    title = "{Current observations are insufficient to confidently associate the binary black hole merger GW190521 with AGN J124942.3 + 344929}",
    eprint = "2009.12346",
    archivePrefix = "arXiv",
    primaryClass = "astro-ph.HE",
    doi = "10.1088/1361-6382/ac33bb",
    journal = "Class. Quant. Grav.",
    volume = "38",
    number = "23",
    pages = "235004",
    year = "2021"
}

@article{Gayathri:2020mra,
    author = "Gayathri, V. and Healy, J. and Lange, J. and O'Brien, B. and Szczepanczyk, M. and Bartos, I. and Campanelli, M. and Klimenko, S. and Lousto, C. O. and O'Shaughnessy, R.",
    title = "{Measuring the Hubble Constant with GW190521 as an Eccentric black hole Merger and Its Potential Electromagnetic Counterpart}",
    eprint = "2009.14247",
    archivePrefix = "arXiv",
    primaryClass = "astro-ph.HE",
    doi = "10.3847/2041-8213/abe388",
    journal = "Astrophys. J. Lett.",
    volume = "908",
    number = "2",
    pages = "L34",
    year = "2021"
}

@article{Mukherjee:2020kki,
    author = "Mukherjee, Suvodip and Ghosh, Archisman and Graham, Matthew J. and Karathanasis, Christos and Kasliwal, Mansi M. and Maga{\~n}a Hernandez, Ignacio and Nissanke, Samaya M. and Silvestri, Alessandra and Wandelt, Benjamin D.",
    title = "{First measurement of the Hubble parameter from bright binary black hole GW190521}",
    eprint = "2009.14199",
    archivePrefix = "arXiv",
    primaryClass = "astro-ph.CO",
    month = "9",
    year = "2020"
}

@article{Graham:2020gwr,
    author = "Graham, M. J. and others",
    title = "{Candidate Electromagnetic Counterpart to the Binary Black Hole Merger Gravitational Wave Event S190521g}",
    eprint = "2006.14122",
    archivePrefix = "arXiv",
    primaryClass = "astro-ph.HE",
    doi = "10.1103/PhysRevLett.124.251102",
    journal = "Phys. Rev. Lett.",
    volume = "124",
    number = "25",
    pages = "251102",
    year = "2020"
}

@article{DelPozzo:2011vcw,
    author = "Del Pozzo, Walter",
    title = "{Inference of the cosmological parameters from gravitational waves: application to second generation interferometers}",
    eprint = "1108.1317",
    archivePrefix = "arXiv",
    primaryClass = "astro-ph.CO",
    doi = "10.1103/PhysRevD.86.043011",
    journal = "Phys. Rev. D",
    volume = "86",
    pages = "043011",
    year = "2012"
}

@article{Chen:2017rfc,
    author = "Chen, Hsin-Yu and Fishbach, Maya and Holz, Daniel E.",
    title = "{A two per cent Hubble constant measurement from standard sirens within five years}",
    eprint = "1712.06531",
    archivePrefix = "arXiv",
    primaryClass = "astro-ph.CO",
    doi = "10.1038/s41586-018-0606-0",
    journal = "Nature",
    volume = "562",
    number = "7728",
    pages = "545--547",
    year = "2018"
}

@article{LIGOScientific:2018gmd,
    author = "Fishbach, M. and others",
    collaboration = "LIGO Scientific, Virgo",
    title = "{A Standard Siren Measurement of the Hubble Constant from GW170817 without the Electromagnetic Counterpart}",
    eprint = "1807.05667",
    archivePrefix = "arXiv",
    primaryClass = "astro-ph.CO",
    reportNumber = "LIGO-P1800192",
    doi = "10.3847/2041-8213/aaf96e",
    journal = "Astrophys. J. Lett.",
    volume = "871",
    number = "1",
    pages = "L13",
    year = "2019"
}

@article{Gray:2019ksv,
    author = "Gray, Rachel and others",
    title = "{Cosmological inference using gravitational wave standard sirens: A mock data analysis}",
    eprint = "1908.06050",
    archivePrefix = "arXiv",
    primaryClass = "gr-qc",
    reportNumber = "LIGO-P1900017",
    doi = "10.1103/PhysRevD.101.122001",
    journal = "Phys. Rev. D",
    volume = "101",
    number = "12",
    pages = "122001",
    year = "2020"
}

@article{Gray:2021sew,
    author = "Gray, Rachel and Messenger, Chris and Veitch, John",
    title = "{A pixelated approach to galaxy catalogue incompleteness: improving the dark siren measurement of the Hubble constant}",
    eprint = "2111.04629",
    archivePrefix = "arXiv",
    primaryClass = "astro-ph.CO",
    doi = "10.1093/mnras/stac366",
    journal = "Mon. Not. Roy. Astron. Soc.",
    volume = "512",
    number = "1",
    pages = "1127--1140",
    year = "2022"
}

@article{Oguri:2016dgk,
    author = "Oguri, Masamune",
    title = "{Measuring the distance-redshift relation with the cross-correlation of gravitational wave standard sirens and galaxies}",
    eprint = "1603.02356",
    archivePrefix = "arXiv",
    primaryClass = "astro-ph.CO",
    doi = "10.1103/PhysRevD.93.083511",
    journal = "Phys. Rev. D",
    volume = "93",
    number = "8",
    pages = "083511",
    year = "2016"
}

@article{Mukherjee:2019wcg,
    author = "Mukherjee, Suvodip and Wandelt, Benjamin D. and Silk, Joseph",
    title = "{Probing the theory of gravity with gravitational lensing of gravitational waves and galaxy surveys}",
    eprint = "1908.08951",
    archivePrefix = "arXiv",
    primaryClass = "astro-ph.CO",
    doi = "10.1093/mnras/staa827",
    journal = "Mon. Not. Roy. Astron. Soc.",
    volume = "494",
    number = "2",
    pages = "1956--1970",
    year = "2020"
}

@article{Mukherjee:2022afz,
    author = "Mukherjee, Suvodip and Krolewski, Alex and Wandelt, Benjamin D. and Silk, Joseph",
    title = "{Cross-correlating dark sirens and galaxies: constraints on $H_0$ from GWTC-3 of LIGO-Virgo-KAGRA}",
    eprint = "2203.03643",
    archivePrefix = "arXiv",
    primaryClass = "astro-ph.CO",
    doi = "10.3847/1538-4357/ad7d90",
    journal = "Astrophys. J.",
    volume = "975",
    number = "2",
    pages = "189",
    year = "2024"
}

@article{Bera:2020jhx,
    author = "Bera, Sayantani and Rana, Divya and More, Surhud and Bose, Sukanta",
    title = "{Incompleteness Matters Not: Inference of $H_0$ from Binary Black Hole{\textendash}Galaxy Cross-correlations}",
    eprint = "2007.04271",
    archivePrefix = "arXiv",
    primaryClass = "astro-ph.CO",
    reportNumber = "LIGO-P2000239-v2",
    doi = "10.3847/1538-4357/abb4e0",
    journal = "Astrophys. J.",
    volume = "902",
    number = "1",
    pages = "79",
    year = "2020"
}

@article{Mukherjee:2021rtw,
    author = "Mukherjee, Suvodip",
    title = "{The redshift dependence of black hole mass distribution: is it reliable for standard sirens cosmology?}",
    eprint = "2112.10256",
    archivePrefix = "arXiv",
    primaryClass = "astro-ph.CO",
    doi = "10.1093/mnras/stac2152",
    journal = "Mon. Not. Roy. Astron. Soc.",
    volume = "515",
    number = "4",
    pages = "5495--5505",
    year = "2022"
}

@article{Chernoff:1993th,
    author = "Chernoff, David F. and Finn, Lee Samuel",
    title = "{Gravitational radiation, inspiraling binaries, and cosmology}",
    eprint = "gr-qc/9304020",
    archivePrefix = "arXiv",
    reportNumber = "NU-GR-4",
    doi = "10.1086/186898",
    journal = "Astrophys. J. Lett.",
    volume = "411",
    pages = "L5--L8",
    year = "1993"
}

@article{Taylor:2011fs,
    author = "Taylor, Stephen R. and Gair, Jonathan R. and Mandel, Ilya",
    title = "{Hubble without the Hubble: Cosmology using advanced gravitational-wave detectors alone}",
    eprint = "1108.5161",
    archivePrefix = "arXiv",
    primaryClass = "gr-qc",
    doi = "10.1103/PhysRevD.85.023535",
    journal = "Phys. Rev. D",
    volume = "85",
    pages = "023535",
    year = "2012"
}

@article{Farr:2019twy,
    author = "Farr, Will M. and Fishbach, Maya and Ye, Jiani and Holz, Daniel",
    title = "{A Future Percent-Level Measurement of the Hubble Expansion at Redshift 0.8 With Advanced LIGO}",
    eprint = "1908.09084",
    archivePrefix = "arXiv",
    primaryClass = "astro-ph.CO",
    reportNumber = "LIGO P1900252",
    doi = "10.3847/2041-8213/ab4284",
    journal = "Astrophys. J. Lett.",
    volume = "883",
    number = "2",
    pages = "L42",
    year = "2019"
}

@article{Mastrogiovanni:2021wsd,
    author = "Mastrogiovanni, S. and Leyde, K. and Karathanasis, C. and Chassande-Mottin, E. and Steer, D. A. and Gair, J. and Ghosh, A. and Gray, R. and Mukherjee, S. and Rinaldi, S.",
    title = "{On the importance of source population models for gravitational-wave cosmology}",
    eprint = "2103.14663",
    archivePrefix = "arXiv",
    primaryClass = "gr-qc",
    doi = "10.1103/PhysRevD.104.062009",
    journal = "Phys. Rev. D",
    volume = "104",
    number = "6",
    pages = "062009",
    year = "2021"
}

@article{LIGOScientific:2025jau,
    author = "Abac, A. G. and others",
    collaboration = "LIGO Scientific, VIRGO, KAGRA",
    title = "{GWTC-4.0: Constraints on the Cosmic Expansion Rate and Modified Gravitational-wave Propagation}",
    eprint = "2509.04348",
    archivePrefix = "arXiv",
    primaryClass = "astro-ph.CO",
    reportNumber = "LIGO-P2400152",
    month = "9",
    year = "2025"
}

@article{Karathanasis:2022rtr,
    author = "Karathanasis, Christos and Mukherjee, Suvodip and Mastrogiovanni, Simone",
    title = "{Binary black holes population and cosmology in new lights: signature of PISN mass and formation channel in GWTC-3}",
    eprint = "2204.13495",
    archivePrefix = "arXiv",
    primaryClass = "astro-ph.CO",
    doi = "10.1093/mnras/stad1373",
    journal = "Mon. Not. Roy. Astron. Soc.",
    volume = "523",
    number = "3",
    pages = "4539--4555",
    year = "2023"
}

@article{Ghosh:2024cwc,
    author = "Ghosh, Tathagata and Biswas, Bhaskar and Bose, Sukanta and Kapadia, Shasvath J.",
    title = "{Joint Inference of Population, Cosmology, and Neutron Star Equation of State from Gravitational Waves of Dark Binary Neutron Stars}",
    eprint = "2407.16669",
    archivePrefix = "arXiv",
    primaryClass = "gr-qc",
    reportNumber = "LIGO-P2400303",
    doi = "10.3847/1538-4365/ae0472",
    journal = "Astrophys. J. Suppl.",
    volume = "281",
    number = "1",
    pages = "11",
    year = "2025"
}

@article{Ghosh:2022muc,
    author = "Ghosh, Tathagata and Biswas, Bhaskar and Bose, Sukanta",
    title = "{Simultaneous inference of neutron star equation of state and the Hubble constant with a population of merging neutron stars}",
    eprint = "2203.11756",
    archivePrefix = "arXiv",
    primaryClass = "astro-ph.CO",
    reportNumber = "LIGO-P2200017",
    doi = "10.1103/PhysRevD.106.123529",
    journal = "Phys. Rev. D",
    volume = "106",
    number = "12",
    pages = "123529",
    year = "2022"
}

@article{Messenger:2011gi,
    author = "Messenger, C. and Read, J.",
    title = "{Measuring a cosmological distance-redshift relationship using only gravitational wave observations of binary neutron star coalescences}",
    eprint = "1107.5725",
    archivePrefix = "arXiv",
    primaryClass = "gr-qc",
    doi = "10.1103/PhysRevLett.108.091101",
    journal = "Phys. Rev. Lett.",
    volume = "108",
    pages = "091101",
    year = "2012"
}

@article{LIGOScientific:2017ycc,
    author = "Abbott, B. P. and others",
    collaboration = "LIGO Scientific, Virgo",
    title = "{GW170814: A Three-Detector Observation of Gravitational Waves from a Binary Black Hole Coalescence}",
    eprint = "1709.09660",
    archivePrefix = "arXiv",
    primaryClass = "gr-qc",
    doi = "10.1103/PhysRevLett.119.141101",
    journal = "Phys. Rev. Lett.",
    volume = "119",
    number = "14",
    pages = "141101",
    year = "2017"
}

@article{DES:2019ccw,
    author = "Soares-Santos, M. and others",
    collaboration = "DES, LIGO Scientific, Virgo",
    title = "{First Measurement of the Hubble Constant from a Dark Standard Siren using the Dark Energy Survey Galaxies and the LIGO/Virgo Binary{\textendash}Black-hole Merger GW170814}",
    eprint = "1901.01540",
    archivePrefix = "arXiv",
    primaryClass = "astro-ph.CO",
    reportNumber = "FERMILAB-PUB-18-629-AE",
    doi = "10.3847/2041-8213/ab14f1",
    journal = "Astrophys. J. Lett.",
    volume = "876",
    number = "1",
    pages = "L7",
    year = "2019"
}

@article{Mozzon:2021wam,
    author = "Mozzon, Simone and Ashton, Gregory and Nuttall, Laura K. and Williamson, Andrew R.",
    title = "{Does nonstationary noise in LIGO and Virgo affect the estimation of H0?}",
    eprint = "2110.11731",
    archivePrefix = "arXiv",
    primaryClass = "astro-ph.CO",
    doi = "10.1103/PhysRevD.106.043504",
    journal = "Phys. Rev. D",
    volume = "106",
    number = "4",
    pages = "043504",
    year = "2022"
}

@article{Veitch:2008ur,
    author = "Veitch, John and Vecchio, Alberto",
    title = "{A Bayesian approach to the follow-up of candidate gravitational wave signals}",
    eprint = "0801.4313",
    archivePrefix = "arXiv",
    primaryClass = "gr-qc",
    reportNumber = "LIGO-P080005-00-Z",
    doi = "10.1103/PhysRevD.78.022001",
    journal = "Phys. Rev. D",
    volume = "78",
    pages = "022001",
    year = "2008"
}

@article{Umstatter:2007xfq,
    author = "Umstatter, Richard and Tinto, Massimo",
    title = "{Bayesian comparison of Post-Newtonian approximations of gravitational wave chirp signals}",
    eprint = "0712.1030",
    archivePrefix = "arXiv",
    primaryClass = "gr-qc",
    doi = "10.1103/PhysRevD.77.082002",
    journal = "Phys. Rev. D",
    volume = "77",
    pages = "082002",
    year = "2008"
}

@article{Apostolatos:1995pj,
    author = "Apostolatos, T. A.",
    title = "{Search templates for gravitational waves from precessing, inspiraling binaries}",
    doi = "10.1103/PhysRevD.52.605",
    journal = "Phys. Rev. D",
    volume = "52",
    pages = "605--620",
    year = "1995"
}

@article{Lindblom:2008cm,
    author = "Lindblom, Lee and Owen, Benjamin J. and Brown, Duncan A.",
    title = "{Model Waveform Accuracy Standards for Gravitational Wave Data Analysis}",
    eprint = "0809.3844",
    archivePrefix = "arXiv",
    primaryClass = "gr-qc",
    doi = "10.1103/PhysRevD.78.124020",
    journal = "Phys. Rev. D",
    volume = "78",
    pages = "124020",
    year = "2008"
}

@article{Damour:2010zb,
    author = "Damour, Thibault and Nagar, Alessandro and Trias, Miguel",
    title = "{Accuracy and effectualness of closed-form, frequency-domain waveforms for non-spinning black hole binaries}",
    eprint = "1009.5998",
    archivePrefix = "arXiv",
    primaryClass = "gr-qc",
    reportNumber = "LIGO-P1000099-V3",
    doi = "10.1103/PhysRevD.83.024006",
    journal = "Phys. Rev. D",
    volume = "83",
    pages = "024006",
    year = "2011"
}

@article{Varma:2018mmi,
    author = "Varma, Vijay and Field, Scott E. and Scheel, Mark A. and Blackman, Jonathan and Kidder, Lawrence E. and Pfeiffer, Harald P.",
    title = "{Surrogate model of hybridized numerical relativity binary black hole waveforms}",
    eprint = "1812.07865",
    archivePrefix = "arXiv",
    primaryClass = "gr-qc",
    doi = "10.1103/PhysRevD.99.064045",
    journal = "Phys. Rev. D",
    volume = "99",
    number = "6",
    pages = "064045",
    year = "2019"
}

@article{Yoo:2023spi,
    author = "Yoo, Jooheon and others",
    title = "{Numerical relativity surrogate model with memory effects and post-Newtonian hybridization}",
    eprint = "2306.03148",
    archivePrefix = "arXiv",
    primaryClass = "gr-qc",
    doi = "10.1103/PhysRevD.108.064027",
    journal = "Phys. Rev. D",
    volume = "108",
    number = "6",
    pages = "064027",
    year = "2023"
}

@article{Scheel:2025jct,
    author = "Scheel, Mark A. and others",
    title = "{The SXS collaboration{\textquoteright}s third catalog of binary black hole simulations}",
    eprint = "2505.13378",
    archivePrefix = "arXiv",
    primaryClass = "gr-qc",
    doi = "10.1088/1361-6382/adfd34",
    journal = "Class. Quant. Grav.",
    volume = "42",
    number = "19",
    pages = "195017",
    year = "2025"
}

@article{Pompili:2023tna,
    author = "Pompili, Lorenzo and others",
    title = "{Laying the foundation of the effective-one-body waveform models SEOBNRv5: Improved accuracy and efficiency for spinning nonprecessing binary black holes}",
    eprint = "2303.18039",
    archivePrefix = "arXiv",
    primaryClass = "gr-qc",
    doi = "10.1103/PhysRevD.108.124035",
    journal = "Phys. Rev. D",
    volume = "108",
    number = "12",
    pages = "124035",
    year = "2023"
}

@article{Ramos-Buades:2023ehm,
    author = "Ramos-Buades, Antoni and Buonanno, Alessandra and Estell{\'e}s, H{\'e}ctor and Khalil, Mohammed and Mihaylov, Deyan P. and Ossokine, Serguei and Pompili, Lorenzo and Shiferaw, Mahlet",
    title = "{Next generation of accurate and efficient multipolar precessing-spin effective-one-body waveforms for binary black holes}",
    eprint = "2303.18046",
    archivePrefix = "arXiv",
    primaryClass = "gr-qc",
    doi = "10.1103/PhysRevD.108.124037",
    journal = "Phys. Rev. D",
    volume = "108",
    number = "12",
    pages = "124037",
    year = "2023"
}

@article{Nagar:2023zxh,
    author = "Nagar, Alessandro and Rettegno, Piero and Gamba, Rossella and Albanesi, Simone and Albertini, Angelica and Bernuzzi, Sebastiano",
    title = "{Analytic systematics in next generation of effective-one-body gravitational waveform models for future observations}",
    eprint = "2304.09662",
    archivePrefix = "arXiv",
    primaryClass = "gr-qc",
    doi = "10.1103/PhysRevD.108.124018",
    journal = "Phys. Rev. D",
    volume = "108",
    number = "12",
    pages = "124018",
    year = "2023"
}

@article{Riemenschneider:2021ppj,
    author = "Riemenschneider, Gunnar and Rettegno, Piero and Breschi, Matteo and Albertini, Angelica and Gamba, Rossella and Bernuzzi, Sebastiano and Nagar, Alessandro",
    title = "{Assessment of consistent next-to-quasicircular corrections and postadiabatic approximation in effective-one-body multipolar waveforms for binary black hole coalescences}",
    eprint = "2104.07533",
    archivePrefix = "arXiv",
    primaryClass = "gr-qc",
    doi = "10.1103/PhysRevD.104.104045",
    journal = "Phys. Rev. D",
    volume = "104",
    number = "10",
    pages = "104045",
    year = "2021"
}

@article{Hamilton:2025xru,
    author = "Hamilton, Eleanor and others",
    title = "{PhenomXPNR: An improved gravitational wave model linking precessing inspirals and NR-calibrated merger-ringdown}",
    eprint = "2507.02604",
    archivePrefix = "arXiv",
    primaryClass = "gr-qc",
    month = "7",
    year = "2025"
}

@article{Colleoni:2024knd,
    author = "Colleoni, Marta and Vidal, Felip A. Ramis and Garc{\'\i}a-Quir{\'o}s, Cecilio and Ak{\c{c}}ay, Sarp and Bera, Sayantani",
    title = "{Fast frequency-domain gravitational waveforms for precessing binaries with a new twist}",
    eprint = "2412.16721",
    archivePrefix = "arXiv",
    primaryClass = "gr-qc",
    doi = "10.1103/PhysRevD.111.104019",
    journal = "Phys. Rev. D",
    volume = "111",
    number = "10",
    pages = "104019",
    year = "2025"
}

@article{Thompson:2023ase,
    author = "Thompson, Jonathan E. and Hamilton, Eleanor and London, Lionel and Ghosh, Shrobana and Kolitsidou, Panagiota and Hoy, Charlie and Hannam, Mark",
    title = "{PhenomXO4a: a phenomenological gravitational-wave model for precessing black-hole binaries with higher multipoles and asymmetries}",
    eprint = "2312.10025",
    archivePrefix = "arXiv",
    primaryClass = "gr-qc",
    reportNumber = "LIGO-P2300437",
    doi = "10.1103/PhysRevD.109.063012",
    journal = "Phys. Rev. D",
    volume = "109",
    number = "6",
    pages = "063012",
    year = "2024"
}

@article{Pratten:2020ceb,
    author = "Pratten, Geraint and others",
    title = "{Computationally efficient models for the dominant and subdominant harmonic modes of precessing binary black holes}",
    eprint = "2004.06503",
    archivePrefix = "arXiv",
    primaryClass = "gr-qc",
    doi = "10.1103/PhysRevD.103.104056",
    journal = "Phys. Rev. D",
    volume = "103",
    number = "10",
    pages = "104056",
    year = "2021"
}

@article{Chatziioannou:2017tdw,
    author = "Chatziioannou, Katerina and Klein, Antoine and Yunes, Nicol{\'a}s and Cornish, Neil",
    title = "{Constructing Gravitational Waves from Generic Spin-Precessing Compact Binary Inspirals}",
    eprint = "1703.03967",
    archivePrefix = "arXiv",
    primaryClass = "gr-qc",
    doi = "10.1103/PhysRevD.95.104004",
    journal = "Phys. Rev. D",
    volume = "95",
    number = "10",
    pages = "104004",
    year = "2017"
}

@article{Storn:1997uea,
    author = "Storn, Rainer and Price, Kenneth",
    title = "{Differential Evolution {\textendash} A Simple and Efficient Heuristic for global Optimization over Continuous Spaces}",
    doi = "10.1023/A:1008202821328",
    journal = "J. Global Optim.",
    volume = "11",
    number = "4",
    pages = "341--359",
    year = "1997"
}

@article{Gao:2012guu,
    author = "Gao, Fuchang and Han, Lixing",
    title = "{Implementing the Nelder-Mead simplex algorithm with~adaptive parameters}",
    doi = "10.1007/s10589-010-9329-3",
    journal = "Comput. Optim. Appl.",
    volume = "51",
    number = "1",
    pages = "259--277",
    year = "2012"
}

@ARTICLE{2020SciPy-NMeth,
  author  = {Virtanen, Pauli and Gommers, Ralf and Oliphant, Travis E. and
            Haberland, Matt and Reddy, Tyler and Cournapeau, David and
            Burovski, Evgeni and Peterson, Pearu and Weckesser, Warren and
            Bright, Jonathan and {van der Walt}, St{\'e}fan J. and
            Brett, Matthew and Wilson, Joshua and Millman, K. Jarrod and
            Mayorov, Nikolay and Nelson, Andrew R. J. and Jones, Eric and
            Kern, Robert and Larson, Eric and Carey, C J and
            Polat, {\.I}lhan and Feng, Yu and Moore, Eric W. and
            {VanderPlas}, Jake and Laxalde, Denis and Perktold, Josef and
            Cimrman, Robert and Henriksen, Ian and Quintero, E. A. and
            Harris, Charles R. and Archibald, Anne M. and
            Ribeiro, Ant{\^o}nio H. and Pedregosa, Fabian and
            {van Mulbregt}, Paul and {SciPy 1.0 Contributors}},
  title   = {{{SciPy} 1.0: Fundamental Algorithms for Scientific
            Computing in Python}},
  journal = {Nature Methods},
  year    = {2020},
  volume  = {17},
  pages   = {261--272},
  adsurl  = {https://rdcu.be/b08Wh},
  doi     = {10.1038/s41592-019-0686-2},
}

@article{Planck:2018vyg,
    author = "Aghanim, N. and others",
    collaboration = "Planck",
    title = "{Planck 2018 results. VI. Cosmological parameters}",
    eprint = "1807.06209",
    archivePrefix = "arXiv",
    primaryClass = "astro-ph.CO",
    doi = "10.1051/0004-6361/201833910",
    journal = "Astron. Astrophys.",
    volume = "641",
    pages = "A6",
    year = "2020",
    note = "[Erratum: Astron.Astrophys. 652, C4 (2021)]"
}

@article{Leong:2025qiw,
    author = "Leong, Samson H. W. and Calder{\'o}n Bustillo, Juan",
    title = "{Kick {\&} spin: new probes for multi-messenger black-hole mergers in AGNs}",
    eprint = "2512.08382",
    archivePrefix = "arXiv",
    primaryClass = "astro-ph.HE",
    reportNumber = "LIGO DCC -- P2500748",
    month = "12",
    year = "2025"
}

@article{Ashton:2018jfp,
    author = "Ashton, Gregory and others",
    title = "{BILBY: A user-friendly Bayesian inference library for gravitational-wave astronomy}",
    eprint = "1811.02042",
    archivePrefix = "arXiv",
    primaryClass = "astro-ph.IM",
    doi = "10.3847/1538-4365/ab06fc",
    journal = "Astrophys. J. Suppl.",
    volume = "241",
    number = "2",
    pages = "27",
    year = "2019"
}

@article{Speagle:2019ivv,
    author = "Speagle, Joshua S.",
    title = "{dynesty: a dynamic nested sampling package for estimating Bayesian posteriors and evidences}",
    eprint = "1904.02180",
    archivePrefix = "arXiv",
    primaryClass = "astro-ph.IM",
    doi = "10.1093/mnras/staa278",
    journal = "Mon. Not. Roy. Astron. Soc.",
    volume = "493",
    number = "3",
    pages = "3132--3158",
    year = "2020"
}

@article{Talbot:2018cva,
    author = "Talbot, Colm and Thrane, Eric",
    title = "{Measuring the binary black hole mass spectrum with an astrophysically motivated parameterization}",
    eprint = "1801.02699",
    archivePrefix = "arXiv",
    primaryClass = "astro-ph.HE",
    doi = "10.3847/1538-4357/aab34c",
    journal = "Astrophys. J.",
    volume = "856",
    number = "2",
    pages = "173",
    year = "2018"
}

@article{Talbot:2017yur,
    author = "Talbot, Colm and Thrane, Eric",
    title = "{Determining the population properties of spinning black holes}",
    eprint = "1704.08370",
    archivePrefix = "arXiv",
    primaryClass = "astro-ph.HE",
    doi = "10.1103/PhysRevD.96.023012",
    journal = "Phys. Rev. D",
    volume = "96",
    number = "2",
    pages = "023012",
    year = "2017"
}

@article{Mapelli:2020vfa,
    author = "Mapelli, Michela",
    title = "{Binary Black Hole Mergers: Formation and Populations}",
    eprint = "2105.12455",
    archivePrefix = "arXiv",
    primaryClass = "astro-ph.HE",
    doi = "10.3389/fspas.2020.00038",
    journal = "Front. Astron. Space Sci.",
    volume = "7",
    pages = "38",
    year = "2020"
}

@article{Madau:2014bja,
    author = "Madau, Piero and Dickinson, Mark",
    title = "{Cosmic Star Formation History}",
    eprint = "1403.0007",
    archivePrefix = "arXiv",
    primaryClass = "astro-ph.CO",
    doi = "10.1146/annurev-astro-081811-125615",
    journal = "Ann. Rev. Astron. Astrophys.",
    volume = "52",
    pages = "415--486",
    year = "2014"
}

@article{Mandel:2018hfr,
    author = "Mandel, Ilya and Farmer, Alison",
    title = "{Merging stellar-mass binary black holes}",
    eprint = "1806.05820",
    archivePrefix = "arXiv",
    primaryClass = "astro-ph.HE",
    doi = "10.1016/j.physrep.2022.01.003",
    journal = "Phys. Rept.",
    volume = "955",
    pages = "1--24",
    year = "2022"
}

@article{Ashton:2017ykh,
    author = "Ashton, G. and Burns, E. and Canton, T. Dal and Dent, T. and Eggenstein, H. -B and Nielsen, A. B. and Prix, R. and Was, M. and Zhu, S. J.",
    title = "{Coincident detection significance in multimessenger astronomy}",
    eprint = "1712.05392",
    archivePrefix = "arXiv",
    primaryClass = "astro-ph.HE",
    doi = "10.3847/1538-4357/aabfd2",
    journal = "Astrophys. J.",
    volume = "860",
    number = "1",
    pages = "6",
    year = "2018"
}

@article{Cotesta:2018fcv,
    author = "Cotesta, Roberto and Buonanno, Alessandra and Boh{\'e}, Alejandro and Taracchini, Andrea and Hinder, Ian and Ossokine, Serguei",
    title = "{Enriching the Symphony of Gravitational Waves from Binary Black Holes by Tuning Higher Harmonics}",
    eprint = "1803.10701",
    archivePrefix = "arXiv",
    primaryClass = "gr-qc",
    doi = "10.1103/PhysRevD.98.084028",
    journal = "Phys. Rev. D",
    volume = "98",
    number = "8",
    pages = "084028",
    year = "2018"
}

\appendix
\twocolumngrid

%%%%%%%%%%%%%%%%%%%%%%%%%%%%%%%%%%%%%%%%%
\section{Origin of biases due to waveform systematics}
\label{app:origin}
%%%%%%%%%%%%%%%%%%%%%%%%%%%%%%%%%%%%%%%%%

Single-event inferences aim to identify waveform parameters \(\boldsymbol{\theta}\) for a chosen model \(M\) such that the template \(h(\boldsymbol{\theta} \mid M)\) adequately represent the true signal \(s(\boldsymbol{\lambda})\), leaving residuals \(r = d - h(\boldsymbol{\theta}\mid M)\) that are consistent with detector noise \(n\) for data \(d\)~\citep{Umstatter:2007xfq, Veitch:2008ur}. Ideally, we expect \(\boldsymbol{\lambda} \approx \boldsymbol{\theta}\) and therefore expect \(s(\boldsymbol{\lambda})-h(\boldsymbol{\theta}\mid M)\rightarrow 0\). In practice, however, imperfect modelling may introduce a discrepancy \(s-h \rightarrow \delta h\) even when the binary parameters are identical~\citep{Lindblom:2008cm}.

Because the posterior depends on the network likelihood of observing the data:
\begin{equation}\label{eq:likelihood}
 \mathcal{L}(d \mid \boldsymbol{\theta}) \propto \prod \exp\left[-\frac{1}{2}\left( d- h \mid d - h \right)\right],
\end{equation}
any systematic discrepancy between \(s\) and \(h\) can affect inference as the stochastic sampler while attempting to minimise \(\left( d- h \mid d - h \right)\) may shift the inferred parameters away from the true ones so as to absorb \(\delta h\)~\footnote{Here \(\left(a \mid b\right)\) represents the noise-weighted inner product:
\[
\left( a \mid b \right)\equiv 4 \Re \sum_k \frac{a^\ast(f_k) b(f_k)}{S_n(f_k)}\,\Delta f 
\]
between two real-valued time series.}. At leading order, the induced parameter bias can be estimated using~\citep{Cutler:2007mi}:
\begin{equation}\label{eq:fisher}
    \Delta \theta_k \simeq \sum_{l} \Gamma^{-1}_{kl} \left(\partial_l h \mid \delta h \right)
\end{equation}
where \(\Gamma_{kl}\) is the Fisher information matrix. This relation emphasises that a waveform discrepancy biases inferred parameters only when it can be mimicked by adjusting the physical parameters of the model. Discrepancy orthogonal to such parameter-induced variations cannot be absorbed and will only reduce the likelihood.

A complementary view comes from considering the zero-noise limit of the likelihood, when systematic errors are negligible only when the discrepancy satisfies \((\delta h \mid \delta h) < \epsilon\), where \(\epsilon\) is a chosen real-valued tolerance. A convenient value of it is one because it ensures that systematic limitations are no worse than statistical limitations. On the other hand, waveform faithfulness~\citep{Apostolatos:1995pj},
\begin{equation}
    \mathcal{F}
    = \max_{\boldsymbol{\Lambda}}
    \frac{ \left( s \mid h(\boldsymbol{\theta}) \right) }
         { \sqrt{ \left( s \mid s \right) \left( h \mid h \right) } }\,,
\end{equation}
which quantifies the agreement between two normalised waveforms after maximisation over a chosen set of parameters \(\boldsymbol{\Lambda} \subset \boldsymbol{\theta}\), is closely related to \((\delta h \mid \delta h)= 2\rho^2(1-\mathcal{F})\). Combining this with Eq.~\eqref{eq:fisher}, the squared displacement of the posterior peak, measured in units of the posterior covariance, satisfies
\begin{equation}
\label{eq:posterior_shift}
    \Delta\theta^k\,\Gamma_{kl}\,\Delta\theta^l \lesssim 2\rho^2\bigl(1-\mathcal{F}\bigr).
\end{equation}

Equation~\eqref{eq:posterior_shift} provides a direct geometric interpretation of waveform unfaithfulness: it quantifies how far the posterior is displaced from the true parameters relative to its own statistical width. Requiring the posterior peak to lie within the nominal \(1\sigma\) region therefore leads to the conservative condition~\citep{Damour:2010zb}
\begin{equation}
\label{eq:criterion}
    2\rho^{2}\bigl(1-\mathcal{F}\bigr) < \epsilon .
\end{equation}
Equation~\eqref{eq:criterion} makes explicit that, in the high-\ac{snr} regime relevant for \ac{xg} detectors, waveform unfaithfulness must scale as \(\bar{\mathcal{F}}=1-\mathcal{F}\sim\mathcal O(\rho^{-2})\) or smaller in order to prevent systematic posterior displacements from exceeding statistical uncertainties.

%%%%%%%%%%%%%%%%%%%%%%%%%%%%%%%%%%%%%%%%%
s\section{Astrophysical population}
    \label{sec:configuration}
%%%%%%%%%%%%%%%%%%%%%%%%%%%%%%%%%%%%%%%%%%

To forecast the impact of systematics on \gds{} cosmology, we construct a synthetic \ac{bbh} population broadly consistent with the distributions inferred in \citet{KAGRA:2021duu}.

\textbf{Redshift model:}
We model the \ac{bbh} redshift distribution using the Madau--Dickinson star-formation rate convolved with a power-law time-delay distribution~\citep{Madau:2014bja}, and use a local merger rate of
\(R_{\rm BBH}=22~\mathrm{Gpc^{-3}\,yr^{-1}}\).
Since our focus is on golden dark sirens, we restrict the population to low redshifts, \(z\leq0.1\).
We generate sources over an observation time of \(T_{\rm obs}=1000~\mathrm{yr}\) for the waveform-systematics study in Sec.~\ref{sec:wsys}, and \(T_{\rm obs}=10~\mathrm{yr}\) for the Bayesian analysis in Sec.~\ref{sec:bayesian}.

\textbf{Mass distribution:}
Component masses are drawn from the \textsc{Power Law}+\textsc{Peak} model introduced in \citet{Talbot:2018cva}.
The primary-mass distribution is a mixture of a doubly truncated power law and a Gaussian component peaked at \(\sim34\,M_\odot\), with a smooth low-mass turn-on.
The mass ratio follows a power-law distribution, ensuring that the secondary mass is always smaller than the primary and shares the same minimum mass of \(\sim5\,M_\odot\) and smoothing scale.

\textbf{Spin distribution:}
Black hole spins are drawn from the \textsc{Default model} of \citet{Talbot:2017yur}.
Spin magnitudes are sampled independently from a beta distribution, while spin-tilt angles are drawn from a mixture of a truncated Gaussian and an isotropic component.
This yields predominantly moderate spin magnitudes with negligible support for large in-plane spins, consistent with expectations from isolated binary formation channels (see \citep{Mandel:2018hfr, Mapelli:2020vfa} and references therein). Given this, we regenerate the spin‑tilt distribution by drawing cosine tilts from a truncated Gaussian in order to explore a wider range of spin orientations, while remaining consistent with the predominantly aligned‑spin nature expected for stellar‑origin binaries.

\textbf{Extrinsic parameters:}
Binary orientation parameters \((\iota,\phi)\) are sampled isotropically, while the polarisation angle \(\psi\) is drawn from a uniform distribution.
For the waveform-systematics study in Sec.~\ref{sec:wsys}, sky locations are assigned isotropically.
For the cosmological analysis in Sec.~\ref{sec:bayesian}, each event is assigned a three-dimensional localization by associating it with galaxies drawn from the MICECATv2.0 catalog~\citep{Fosalba:2013wxa, Fosalba:2013mra, Crocce:2013vda, Carretero:2014ltj, Hoffmann:2014ida}. Since the population is characterized by a non-trivial redshift distribution, the association is performed by assigning the source to a galaxy within $\pm 20$ Mpc from the originally sampled $D_L$, with probability proportional to each galaxy's mass. This is equivalent to sampling the counterpart directly from the catalog, since the thickness of the spherical shell is large enough to enclose the large-scale structure, naturally letting the sources distribute according to the intrinsic galaxy distribution of the catalog. At the same time, the redshift distribution imposed during the population sampling is conserved since the shell thickness is sufficiently small compared to the redshift distribution scale length. Note also that the sky location (\(\alpha\), \(\delta\)) is completely decoupled from the initial sampling.

%%%%%%%%%%%%%%%%%%%%%%%%%%%%%%%%%%%
\subsection{Adapting the population for Sec.~\ref{sec:bayesian}}
%%%%%%%%%%%%%%%%%%%%%%%%%%%%%%%%%%%
When comparing waveform models from a modelling perspective, we find that for the fiducial \ac{bbh} population considered there, \imrxpnr{} provides a better representation of the corresponding \nrhyb{} signals than \seob{}, with the latter having a broader distribution of unfaithfulness, with systematically poorer agreement at lower redshifted total masses \(M_T(1+z)\) (see Fig.~\ref{fig:unfaithfulness-1}). Although allowing the intrinsic parameters to vary can significantly reduce the average unfaithfulness \(\bar{\mathcal{F}}\) (see Fig.~\ref{fig:data-analyst}), we restrict the Bayesian analysis in Sec.~\ref{sec:bayesian} to systems with redshifted chirp mass \(\mathcal{M}\gtrsim15\,M_\odot\).

This choice is motivated primarily by two considerations: 
\begin{inparaenum}[(1)]
    \item \emph{Computational efficiency:} waveform generation for long-duration signals using \seob{} remains computationally expensive, despite the use of post-adiabatic approximations during the inspiral. Since our primary goal is to assess the impact of waveform systematics on \(H_0\) inference—rather than to perform a comprehensive survey of waveform agreement across the full parameter space—we adopt this restriction as a pragmatic compromise.
    \item \emph{Cosmological informativeness:} lower chirp-mass systems yield poorer constraints on \(H_0\) once the contribution from galaxy peculiar velocities is taken into account. For example, systems with \(\mathcal{M}\approx15\,M_\odot\) achieve a relative uncertainty of \(\sigma_{H_0}/H_0 \approx 0.5\%\) for luminosity distances in the range \(D_L\in(400,700)\,\mathrm{Mpc}\), whereas lighter systems are dominated by velocity-induced uncertainties (see Appendix~\ref{app:error_propagation}).
\end{inparaenum}

\begin{figure}
    \centering
    \includegraphics[width=0.45\textwidth]{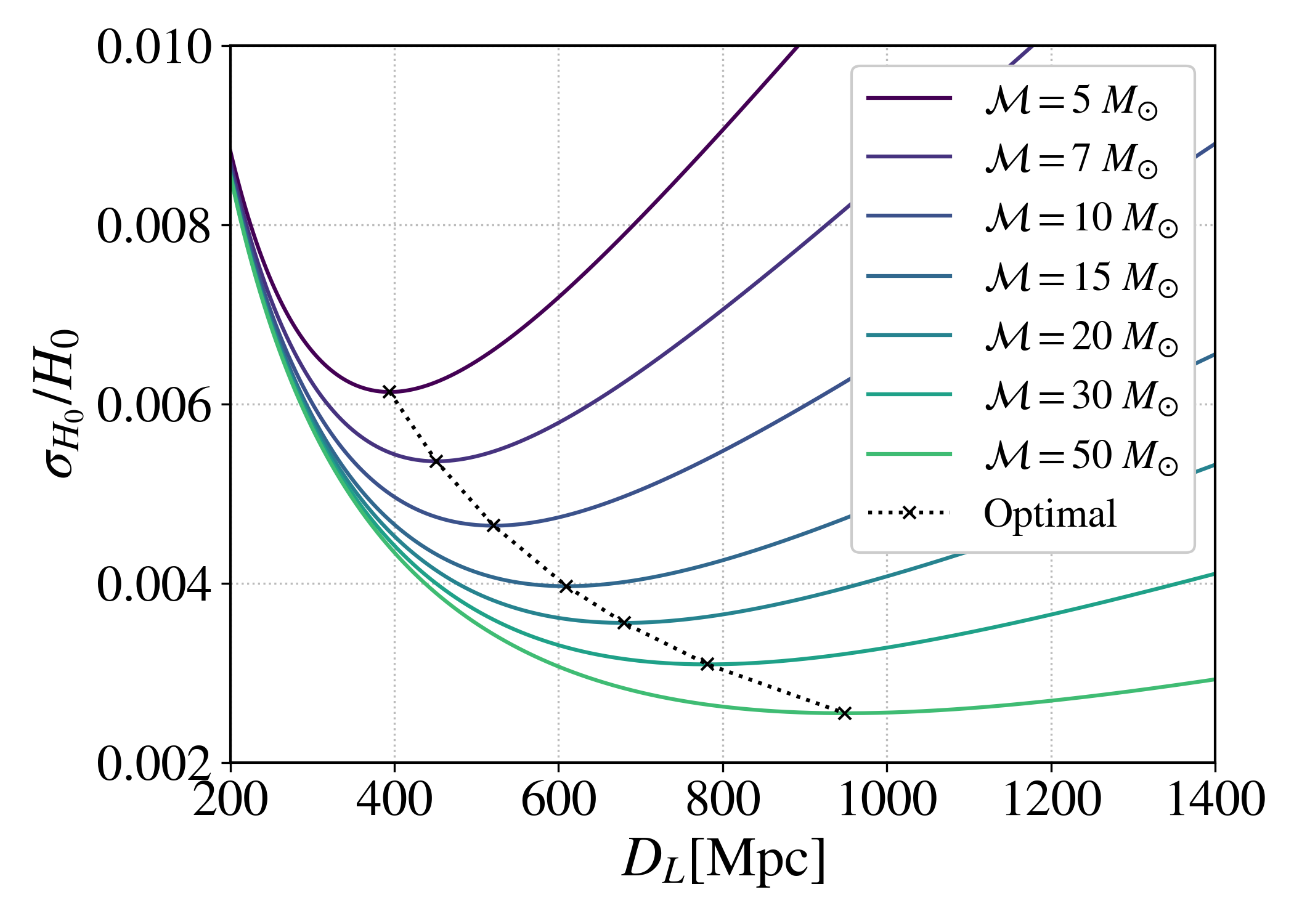}
    \caption{Optimal-source (proxy for \gds{}) errors for different chirp masses $\mathcal{M}$ as a function of the $d_L$. The minima of the error curves are highlighted with a dotted line.}
    \label{fig:H0_error_curves}
\end{figure}

%%%%%%%%%%%%%%%%%%%%%%%%%%%%%%%%%%%%%%%%%
\section{Error propagation for $H_0$}
\label{app:error_propagation}
%%%%%%%%%%%%%%%%%%%%%%%%%%%%%%%%%%%%%%%%%

To have an order-of-magnitude estimate on how errors in single-event parameter inference can impact \(H_0\) measurement, we consider the linearised Hubble law:
\begin{equation}
    D_L = \frac{cz}{H_0},
\end{equation}
which is valid in the low-redshift regime considered here. The relative error on \(H_0\) is
\begin{equation}
\label{eq:H0_relative_error}
    \frac{\sigma_{H_0}}{H_0}
    =
    \sqrt{
        \left(\frac{\sigma_{D_L}}{D_L}\right)^2
        +
        \left(\frac{\sigma_z}{z}\right)^2
    }.
\end{equation}
The error on \(D_L\) arises from \ac{gw} parameter estimation, while the error on \(z\) is dominated by host-galaxy peculiar velocities. We model peculiar velocities as a 3D Gaussian with dispersion \(\sigma_v\simeq200~\mathrm{km\,s^{-1}}\), corresponding to a radial dispersion \(\sigma_{v,r}=\sigma_v/\sqrt{3}\). This gives
\begin{equation}
    \frac{\sigma_z}{z}
    =
    \frac{\sigma_{v,r}}{cz}
    =
    \frac{\sigma_{v,r}}{d_L H_0}
    \equiv
    \frac{\beta}{d_L},
\end{equation}
with \(\beta\simeq1.71~\mathrm{Mpc}\).

The distance error scales inversely with the network optimal \ac{snr},
\begin{equation}\label{eq:skyaverageeffdistance}
    \frac{\sigma_{d_L}}{d_L}
    \propto
    \rho_{\rm opt}^{-1}
    \propto
    D_{\rm eff},
\end{equation}
where the sky-averaged effective distance \(D_{\rm eff}\), assuming that the signal is entirely modelled by the dominant quadrupole mode, is
\begin{equation}
    D_{\rm eff}
    =
    D_L
    \left[
        \frac{1}{2}
        \left(
            \left(\frac{1+\cos^2\iota}{2}\right)^2
            +
            \cos^2\iota
        \right)
    \right]^{-1/2}.
\end{equation}
Averaging over orientations gives
\begin{equation}
    \left\langle \frac{\sigma_{D_L}}{D_L} \right\rangle
    =
    \alpha\, D_L,
\end{equation}
where \(\alpha\) depends on the intrinsic source parameters. We determine \(\alpha\) numerically using the Fisher-matrix code \textsc{GWFish} for optimally oriented \gds{} at \(D_L=400~\mathrm{Mpc}\)~\citep{Dupletsa:2022scg}.

Combining both contributions,
\begin{equation}
    \frac{\sigma_{H_0}}{H_0}
    =
    \sqrt{
        (\alpha D_L)^2
        +
        \left(\frac{\beta}{D_L}\right)^2
    },
\end{equation}
which presents a minimum at \(D_{\rm opt}=\sqrt{\beta/\alpha}\).
We plot in Fig.~\ref{fig:H0_error_curves} the resulting error curves for representative chirp masses $\mathcal{M}$. This shows that not only the informativeness of a golden dark siren increases with chirp mass and fixed $D_L$ as expected, but also that the luminosity distance optimal window widens and shifts towards higher $D_L$, which implies progressively larger available detection volumes. However, the classification of high-$\mathcal{M}$ systems as \gds is limited by the growing number of possible counterparts in the 90\% sky area, despite in principle being highly informative well beyond 1 Gpc. Note that adding a proper motion distribution to the synthetic catalog does not affect the association results presented in Sec. \ref{sec:3DSourcelocalization}, but only introduces a $D_L$-dependent dispersion of the individual $H_0$ posteriors around the true value.

To validate, we apply this prescription to the synthetic \gds{} catalog. For each event \(k\), we draw a redshift perturbation from the peculiar-velocity distribution and infer \(H_0^{(k)}\) with uncertainty given by Eq.~\eqref{eq:H0_relative_error}. Assuming \(\Omega_m\) is known, a Gaussian maximum-likelihood fit recovers the true value of \(H_0\) without bias and with a statistical uncertainty of \(\simeq0.1\%\).

%%%%%%%%%%%%%%%%%%%%%%%%%%%%%%%%%%%%%%%%%%%%%%
\section{Sky location whitening}
\label{app:whitening}
%%%%%%%%%%%%%%%%%%%%%%%%%%%%%%%%%%%%%%%%%%%%%%
\begin{figure*}
    \centering
    \includegraphics[width=1\linewidth]{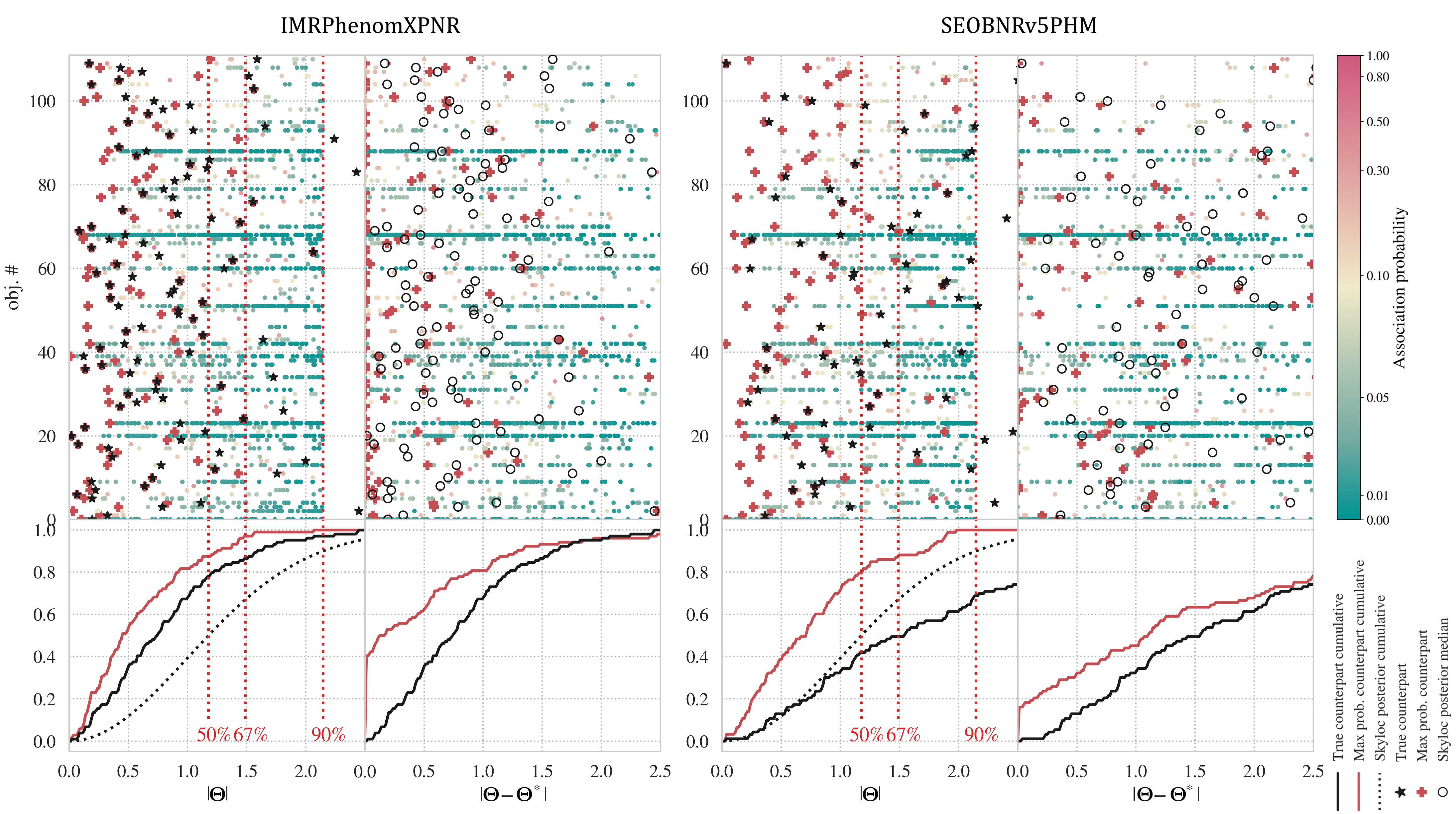}
    \caption{Comparison of sky-localization biases and host-galaxy association for different waveform models. Candidate host galaxies are selected within the 90\% credible sky area and with redshifts consistent with the inferred luminosity distance under a flat prior on $H_0$. Left panels show distances of candidate galaxies from the posterior median in the whitened sky-coordinate plane, with colors indicating association probability; the most probable counterpart is marked by a red plus and the true host by a black star. Cumulative distributions for true and best-match counterparts are shown below, together with the cumulative posterior (dotted curve) and representative confidence levels. Right panels show the complementary distances of candidate galaxies from the true host, with empty markers indicating the posterior medians. The red solid curves denote the cumulative distribution of best-match counterparts, providing a direct measure of host-association success.}
    \label{fig:skyloc_comparison}
\end{figure*}

Here, we describe and motivate the implementation of the whitening procedure in a general setting; The procedure is then specialized to the sky location case. Let $\boldsymbol{\Omega}$ be the vector to be whitened, $\boldsymbol{\Omega}_b$ the best-fit location, and $\mathcal{C}$ the covariance matrix associated with the Gaussian posterior distribution. The whitened vector $\boldsymbol{\Theta}$ is defined as
\begin{equation*}
    \boldsymbol{\Theta} = (\boldsymbol{\Omega}-\boldsymbol{\Omega}_b) \cdot \boldsymbol{E V E}^T
\end{equation*}
where $\boldsymbol{E}$ is the matrix obtained by vertically stacking the eigenvectors $v_i$ of the covariance matrix, and $\boldsymbol{V}$ is the diagonal matrix containing in the $i,i$ position the inverse of the square root of the $i-$th eigenvalue $\lambda_i$ associated with the eigenvector $v_i$:
\begin{equation*}
    \boldsymbol{E} \equiv \left( \begin{array}{ccc} | & | & | \\
 v_1 & ... & v_n \\
 |&|&|\end{array} \right) \hspace{1cm} \boldsymbol{V} \equiv \left( \begin{array}{cccc} \lambda_1^{-\frac{1}{2}} & 0 & ... & 0 \\
 0 & \lambda_2^{-\frac{1}{2}} &... & 0 \\
 ...&...&...&...\\ 0 &0&...& \lambda_n^{-\frac{1}{2}}\end{array} \right) 
\end{equation*}

The linear map represented by the matrix $\boldsymbol{W} \equiv \boldsymbol{EVE}^T$ implements the definition of whitening: if applied to the covariance matrix $\mathcal{C}$ through $\mathcal{C}' = \boldsymbol{W}^T\mathcal{C}\boldsymbol{W}$, by definition the application of $\boldsymbol{E}$ is a basis change that diagonalizes $\mathcal{C}$, the application of $\boldsymbol{V}$ normalizes the matrix in the diagonalized basis and finally $\boldsymbol{E}^T$ de-rotates the normalized matrix back to the original basis. Note that since already $(\boldsymbol{V}^T\boldsymbol{E}^T\mathcal{C}\boldsymbol{EV})_i^j=\delta_i^j$ there is a freedom of choice for the last rotation, namely every $\Tilde{\boldsymbol{W}}= \boldsymbol{EVR}$ with $\boldsymbol{R}$ orthogonal ($\boldsymbol{R}^T=\boldsymbol{R}^{-1}$) represents a valid whitening. The choice of $\boldsymbol{R}=\boldsymbol{E}^T$ takes the name of zero-phase component analysis (ZCA) whitening. It has the property of minimizing the expectancy value of $(\boldsymbol{x} - \boldsymbol{x}\cdot \boldsymbol{W})^2$, where $\boldsymbol{x}$ is drawn from the zero-mean Gaussian distribution associated with the covariance matrix $\mathcal{C}$. This means that the transformed vector retains part of the orientation properties, despite the partial coordinate mixing due to the rescaling of generally different covariance eigenvalues.

In the case of sky location, this procedure is specialized to the reduced covariance matrix $\mathcal{C}_\text{sky}$ associated with the sky location coordinates $\boldsymbol{\Omega}=(\alpha, \cos \delta)$. The whitened coordinate vector is $\boldsymbol{\Theta}=(\theta_\alpha, \theta_\delta)$. The coordinates are labeled with $\alpha$ and $\delta$ since the orientation is loosely conserved; however, we remark that the whitening introduces a partial coordinate mixing. Since the GDS have SNR$\gtrsim 500$, the gaussian approximation of the posterior distribution is generally appropriate. 

%%%%%%%%%%%%%%%%%%%%%%%%%%%%%%%%%%%%%%%%%%%%%%
\section{3D Sky localization of Golden Dark Sirens}
\label{sec:3d}
%%%%%%%%%%%%%%%%%%%%%%%%%%%%%%%%%%%%%%%%%%%%%%

We summarise the sky-localization biases and host-association outcomes in Fig.~\ref{fig:skyloc_comparison}. Galaxy counterparts are restricted to those lying within the 90\% sky credible region and whose redshifts are consistent with the inferred luminosity distance, assuming a flat prior on $H_0$ over $60\text{--}75~\mathrm{km\,s^{-1}\,Mpc^{-1}}$. Sky positions are expressed in whitened coordinates, with the origin at the median of the posterior sky distribution.

To calculate the probability that a given galaxy in the catalog is the host of the \ac{gw} source, we follow the Bayesian formalism introduced in~\citet{Ashton:2017ykh}. The association probability is constructed from the spatial overlap between independent inferences on the common set of source parameters
\(\boldsymbol{\kappa}\), which encodes the position of the source in terms of sky location and luminosity distance.

For each candidate host galaxy indexed by \(\boldsymbol{\kappa}\), we calculate the spatial overlap integral
\begin{equation}
\mathcal{I}_{\boldsymbol{\kappa}}=\frac{p(\boldsymbol{\kappa}_{\rm gal}\mid d_{\rm gw}, C)}{\pi(\boldsymbol{\kappa}_{\rm gal}\mid C)},
\end{equation}
where \(\boldsymbol{\kappa}_{\rm gal}\) denotes the parameters corresponding to the location of the candidate host galaxy. This quantity measures the support from the \ac{gw} data for the source being located at the galaxy's position, relative to the prior alone. We accounted for a flat $H_0$ prior between 60 and 75 km/s/Mpc when comparing the redshift of the electromagnetic counterpart with the GW-measured luminosity distance.

Given a discrete set of candidate host galaxies, the association probability for a particular galaxy is then obtained by normalising the overlap integral,
\begin{equation}
p_{\rm association}(\boldsymbol{\kappa}_{\rm gal})=\frac{\mathcal{I}_{\boldsymbol{\kappa}_{\rm gal}}}{\sum_{\boldsymbol{\kappa}'}\mathcal{I}_{\boldsymbol{\kappa}'}},
\end{equation}
where the sum runs over all galaxies within the localization volume. Under the assumption that all candidate hosts are a priori equally likely, this expression defines a properly normalised posterior probability that the \ac{gw} event is associated with a given galaxy, conditioned on both the \ac{gw} and electromagnetic information.

For each waveform model, the upper-left panel shows the distance of all viable host galaxies from the posterior median in the whitened sky plane. Association probabilities are encoded by color; the most probable host is marked with a red plus, and the true host with a black star. The cumulative distributions of distances for the true and best-match counterparts are shown below, together with the cumulative posterior distribution (dotted line), enabling a direct comparison with the nominal 50\%, 67\%, and 90\% sky credible regions.

For \imrxpnr{} recovery, approximately 80\% of true counterparts lie within the 50\% sky credible region, increasing to 85\% and 95\% within the 67\% and 90\% regions, respectively. These fractions are significantly reduced for \seob{} recovery, dropping to $\sim40\%$, $\sim50\%$, and $\sim70\%$. The right-hand panels show the separation, in whitened coordinates, between each inferred host and the true counterpart. Empty markers denote distances measured from the posterior median. The cumulative distribution of best-match separations provides a direct measure of host-association success: about 40\% of events are correctly associated for \imrxpnr{} recovery, compared to $\sim20\%$ for \seob{}.

Finally, we emphasise that an incorrect host identification does not necessarily lead to a biased measurement of $H_0$, as misidentified galaxies often belong to the same large-scale structure and therefore share similar redshifts.

%%%%%%%%%%%%%%%%%%%%%%%%%%%%%%%%%%%%%%%%%%%%%%
\section{\(H_0\) measurements for \seob{} recovery}
\label{sec:hubble}
%%%%%%%%%%%%%%%%%%%%%%%%%%%%%%%%%%%%%%%%%%%%%%
\begin{figure*}
    \centering
    \includegraphics[width=1\linewidth]{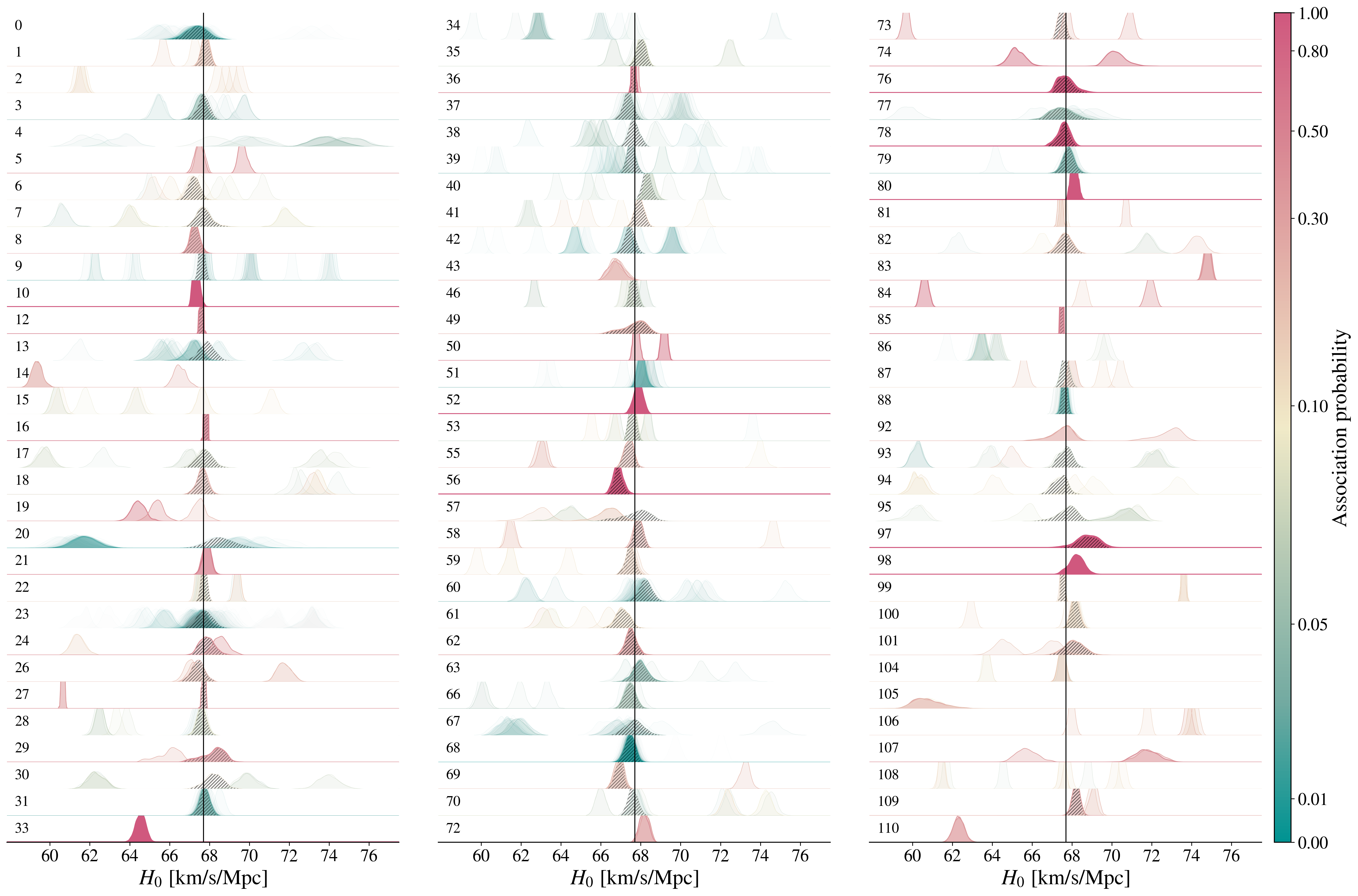}
    \caption{$H_0$ posteriors for \seob{} signal recovery. Same as Fig. \ref{fig:hubble-ridge}. }
    \label{fig:SEOB_H0_posteriors}
\end{figure*}
Analogously to Fig. \ref{fig:hubble-ridge}, we plot the $H_0$ posterior distributions obtained from \seob{}   signal recovery in Fig. \ref{fig:SEOB_H0_posteriors}. The presence of bigger systematics in the reconstructed localization leads to a more frequent galaxy misassociation ( $\approx 30\%$ of the true hosts lie outside the 90\% sky area, corresponding in the figure to the systems which do not display a hatched distribution); this introduces the possibility of observing ``golden dark siren impostors'', namely systems for which only one possible electromagnetic counterpart exists, but it does not correspond to the true source. In the majority of the cases, however (e.g., sys. 10, 52, 98), as observed for \imrxpnr{} signal reconstruction, the impostor belongs to the same group or cluster of the true host, leading to nearly unbiased $H_0$ posteriors. In extremely rare cases (sys. 33), the impostor belongs to a different cluster, leading to a strongly biased $H_0$. More often (e.g. sys. 15, 74, 106, 108), the absence of the true host from the considered counterparts leads to the presence of many poorly informative posteriors which do not show a clear correlation with the true host. Still, similarly to \imrxpnr{} recovery, we find true \gds{} (e.g., sys. 12, 76, 78) as well as ``cumulative'' \gds{}, for which there exist many potential sources with the same luminosity distance and collective probability near unity. 

The effect of the bias induced by waveform discrepancies on the $D_L$ measurement is to slightly shift the individual $H_0$ posteriors (hatched) from the injected value. This effect is more prominent for \seob{} signal recovery compared to \imrxpnr{}, but it is generally less problematic with respect to the galaxy misassociation caused by the biases in sky location.

%%%%%%%%%%%%%%%%%%%%%%%%%%%%%%%%%%%%%%%%%%%%%%%%%%%%%%%%%%%%%%%%%%%%%%%%
\section{Derivation of Eq. 4}
\label{app:calibration}
%%%%%%%%%%%%%%%%%%%%%%%%%%%%%%%%%%%%%%%%%%%%%%%%%%%%%%%%%%%%%%%%%%%%%%%%
% \todo{Sathya do take a look at this math}

The calibrated detector strain is obtained by convolving the raw photodetector output with a frequency‑dependent detector response. Uncertainties in this response lead to both statistical and systematic calibration errors \citep{LIGOScientific:2016xax, LIGOScientific:2017aaj, Sun:2021qcg, Capote:2024rmo}, and the observed strain in the frequency domain is:
\begin{equation}
d_{\mathrm{obs}}(f)
=
d(f)\,[1+\delta A(f\mid t)]\,e^{i\delta\phi(f\mid t)} ,
\end{equation}
where $\delta A$ and $\delta\phi$ denote fractional amplitude and phase offsets.

To estimate the impact of these errors, we consider an idealised scenario in which the \ac{gw} signal is perfectly described by the waveform model, all binary parameters are known exactly, and the data are otherwise noise‑free. Then, any difference between the observed signal and the model waveform,
\begin{equation}
    \delta h \equiv s_{\mathrm{obs}} - s = s\,[1+\delta A(f)]\,e^{i\delta\phi(f)} - s ,
\end{equation}
is entirely due to calibration errors.

Since calibration errors are typically small, \((\delta A,\delta\phi=\mathcal{O}(10^{-2})\) for current detectors, we expand the above equation to linear order,
\begin{equation}
    \delta h(f) \simeq s(f)\,[\delta A(f)+i\,\delta\phi(f)] .
\end{equation}

The corresponding noise‑weighted inner product is
\begin{equation}
\begin{aligned}
(\delta h|\delta h)
&= 4\sum_k
\frac{|s(f_k)|^2}{S_n(f_k)}
\bigl[\delta A^2(f_k)+\delta\phi^2(f_k)\bigr]
\Delta f \\
&= \rho^2\left\langle \delta A^2+\delta\phi^2 \right\rangle ,
\end{aligned}
\end{equation}
where \(S_n(f)\) is the one‑sided noise power spectral density and the angled brackets denote an \ac{snr}‑weighted average.

Requiring calibration errors to be indistinguishable from statistical fluctuations in the likelihood leads to the criterion
\begin{equation}
(\delta h|\delta h) < 1
\quad \Rightarrow \quad
\left\langle \delta A^2+\delta\phi^2 \right\rangle < \rho^{-2} .
\end{equation}

Introducing the per‑frequency \ac{snr} contribution
\begin{equation}
\rho_k^2 = \frac{4|s(f_k)|^2}{S_n(f_k)}\Delta f ,
\end{equation}
this condition becomes
\begin{equation}
    \sum_k \rho_k^2 \bigl[\delta A^2(f_k)+\delta\phi^2(f_k)\bigr] < 1 .
\end{equation}
A sufficient (though conservative) local condition is therefore~\citep{Read:2023hkv}

\begin{equation}
\delta A^2(f_k)+\delta\phi^2(f_k) < \rho_k^{-2}/N = \frac{S_n(f_k)}{4|s(f_k)|^2}\frac{1}{f_\text{max}-f_\text{min}} .
\end{equation}

It is convenient to define a combined amplitude–phase deviation
\begin{equation}
Y(f_k) \equiv \sqrt{\delta A^2(f_k)+\delta\phi^2(f_k)} ,
\end{equation}
leading to a frequency‑dependent tolerance

\begin{equation}
    Y(f_k) < \frac{\sqrt{S_n(f_k)/(f_\text{max}-f_\text{min})}}{2|s(f_k)|}
\end{equation}

In practice, calibration uncertainties are parametrised as smooth functions of frequency using spline bases,
$\delta A(f)=\sum_i a_i B_i(f)$ and $\delta\phi(f)=\sum_j p_j C_j(f)$ \citep{Farr2014Calibration}. Substituting these forms into the indistinguishability criterion yields a quadratic constraint on the spline coefficients,
\begin{equation}
\boldsymbol{a}^T\Gamma^A\boldsymbol{a} + \boldsymbol{p}^T\Gamma^\phi\boldsymbol{p} <1 ,
\end{equation}
where
\begin{equation}
    \Gamma^A_{ii'}=\sum_k \rho_k^2 B_i(f_k)B_{i'}(f_k),
\end{equation}
and analogously for $\Gamma^\phi$. This explicitly shows that calibration parameters whose basis functions overlap with frequency regions that contribute most strongly to the total \ac{snr} must be constrained most tightly \citep{Essick:2022vzl}.

\end{document}